\RequirePackage{fix-cm}
\documentclass[twocolumn, epjc3]{svjour3}
\smartqed  
\RequirePackage[dvipdfmx]{graphicx}
\RequirePackage{units}
\RequirePackage{siunitx}
\RequirePackage{booktabs}
\RequirePackage{subfig}
\RequirePackage{textcomp}
\RequirePackage[centertags]{amsmath}
\RequirePackage{multirow}
\RequirePackage{placeins}
\RequirePackage{mathptmx}      
\RequirePackage[version=4]{mhchem}
\RequirePackage{mwe}  
\RequirePackage{cite}
\RequirePackage{notoccite}
\RequirePackage[colorlinks,linkcolor=blue,anchorcolor=blue,citecolor=blue,urlcolor=blue]{hyperref}
\RequirePackage{bookmark}
\RequirePackage{bmpsize}
\RequirePackage{amssymb}
\RequirePackage[T1]{fontenc}
\journalname{Eur. Phys. J. C}
\begin{document}
\title{Calibration Strategy of the JUNO-TAO Experiment}

\author{
Hangkun Xu \thanksref{a,b} \and
Angel Abusleme \thanksref{c,d} \and
Nikolay V. Anfimov \thanksref{m} \and
Stéphane Callier \thanksref{h} \and
Agustin Campeny \thanksref{c} \and
Guofu Cao \thanksref{a,b} \and
Jun Cao \thanksref{a,b} \and
Cedric Cerna \thanksref{g} \and
Yu Chen \thanksref{e} \and
Alexander Chepurnov \thanksref{o} \and
Yayun Ding \thanksref{a,b} \and
Frederic Druillole \thanksref{g} \and
Andrea Fabbri \thanksref{l} \and
Zhengyong Fei \thanksref{a,b} \and
Maxim Gromov \thanksref{n,o} \and
Miao He \thanksref{a,b} \and
Wei He \thanksref{a,b} \and
Yuanqiang He \thanksref{e} \and
Joseph yk Hor \thanksref{e} \and
Shaojing Hou \thanksref{b} \and
Jianrun Hu \thanksref{a,b} \and
Jun Hu \thanksref{b} \and
Cédric Huss \thanksref{g} \and
Xiaolu Ji \thanksref{a,b} \and
Tao Jiang \thanksref{e} \and
Xiaoshan Jiang \thanksref{a,b} \and
Cécile Jolliet \thanksref{g} \and
Daozheng Li \thanksref{b} \and
Min Li \thanksref{a,b} \and
Ruhui Li \thanksref{a,b} \and
Yichen Li \thanksref{a,b} \and
Caimei Liu \thanksref{b} \and
Mengchao Liu \thanksref{b} \and
Yunzhe Liu \thanksref{a,b} \and
Claudio Lombardo \thanksref{k} \and
Selma Conforti Di Lorenzo \thanksref{h} \and
Peizhi Lu \thanksref{e} \and
Guang Luo \thanksref{e} \and
Mari Stefano M. \thanksref{l} \and
Xiaoyan Ma \thanksref{a,b} \and
Paolo Montini \thanksref{l} \and
Juan Pedro Ochoa-Ricoux \thanksref{c} \and
Yatian Pei \thanksref{a,b} \and
Frédéric Perrot \thanksref{g} \and
Fabrizio Petrucci \thanksref{l} \and
Xiaohui Qian \thanksref{a,b} \and
Abdel Rebii \thanksref{g} \and
Bedřich Roskovec \thanksref{f} \and
Arsenij Rybnikov \thanksref{m} \and
Hans Steiger \thanksref{i,j} \and
Xilei Sun \thanksref{a,b} \and
Pablo Walker \thanksref{c} \and
Derun Wang \thanksref{a,b} \and
Meifen Wang \thanksref{b} \and
Wei Wang \thanksref{e} \and
Wei Wang \thanksref{b} \and
Zhimin Wang \thanksref{a,b} \and
Diru Wu \thanksref{a,b} \and
Xiang Xiao \thanksref{e} \and
Yuguang Xie \thanksref{a,b} \and
Zhangquan Xie \thanksref{a,b} \and
Wenqi Yan \thanksref{a,b} \and
Huan Yang \thanksref{b} \and
Haifeng Yao \thanksref{a,b} \and
Mei Ye \thanksref{a,b} \and
Chengzhuo Yuan \thanksref{a,b} \and
Kirill Zamogilnyi \thanksref{p} \and
Liang Zhan \thanksref{cu,a,b} \and
Jie Zhang \thanksref{a,b} \and
Shuihan Zhang \thanksref{b} \and
Rong Zhao \thanksref{e}
}
\institute{
University of Chinese Academy of Sciences, Beijing, China \label{a} \and
Institute of High Energy Physics, Beijing, China \label{b} \and
Pontificia Universidad Católica de Chile, Santiago, Chile \label{c}\and
Millennium Institute for Subatomic Physics at High-Energy Frontier - SAPHIR, Santiago, Chile \label{d}\and
Sun Yat-sen University, Guangzhou, China \label{e}\and
Faculty of Mathematics and Physics, Charles University, Prague, Czech Republic \label{f}\and
Centre d'Etudes Nucléaires de Bordeaux-Gradignan, Gradignan, France \label{g}\and
Organisation de MicroÉlectronique Générale Avancée, Palaiseau, France \label{h}\and
Physik-Department, Technische Universität München, James-Franck-Str. 1, Garching, Germany \label{i}\and
Cluster of Excellence PRISMA+ and Institute of Physics, Detector Laboratory, Staudingerweg 9, Mainz, Germany \label{j}\and
INFN Catania and Dipartimento di Fisica e Astronomia dell’Università di Catania, Catania, Italy \label{k}\and
Istituto Nazionale di Fisica Nucleare Sezione di Roma Tre, Roma, Italy \label{l}\and
Joint Institute for Nuclear Research, Dubna, Russia \label{m}\and
Moscow State University, Moscow, Russia \label{n}\and
Skobeltsyn Institute of Nuclear Physics, Lomonosov Moscow State University, Moscow, Russia \label{o}\and
Faculty of Physics, Lomonosov Moscow State University, Moscow, Russia \label{p}
}
\thankstext{cu}{e-mail: zhanl@ihep.ac.cn (corresponding author)}

\date{Received: date / Accepted: date}

\maketitle{}

\begin{abstract}
The Taishan Antineutrino Observatory (TAO or JUNO-TAO) is a satellite detector for the Jiangmen Underground Neutrino Observatory (JUNO). Located near the Taishan reactor, TAO independently measures the reactor's antineutrino energy spectrum with unprecedented energy resolution. To achieve this goal, energy response must be well calibrated. Using the Automated Calibration Unit (ACU) and the Cable Loop System (CLS) of TAO, multiple radioactive sources are deployed to various positions in the detector to perform a precise calibration of energy response. The non-linear energy response can be controlled within 0.6\% with different energy points of these radioactive sources. It can be further improved by using $^{12}\rm B$ decay signals produced by cosmic muons. Through the energy non-uniformity calibration, residual non-uniformity is less than 0.2\%. The energy resolution  degradation and energy bias caused by the residual non-uniformity can be controlled within 0.05\% and 0.3\%, respectively. In addition, the stability of other detector parameters, such as the gain of each silicon photo-multiplier, can be monitored with a special ultraviolet LED calibration system.

\end{abstract}

\section{Introduction}
The use of the liquid scintillator (LS) technique to detect reactor antineutrinos endowed great progress in neutrino physics \cite{Cowan:1956rrn,CHOOZ:2002qts,kamland2008precision,dayabay2012observation,reno2012observation,DoubleChooz:2011ymz} and it is also an effective approach which will be used in next-generation experiment such as Jiangmen Underground Neutrino Observatory (JUNO) \cite{juno2016neutrinophys}. 
In LS, reactor antineutrinos are detected via the inverse beta decay (IBD) reaction,
\begin{equation}
    \Bar\nu_e + p \rightarrow e^{+}+n.
\end{equation}
The positron deposits its kinetic energy, then annihilates immediately with the electron and emits mostly two back-to-back gammas. The neutron scatters in the detector until it is thermalized and then captured on a nucleus. Subsequently, the nucleus de-excites with the emission of gamma rays. Since neutron capture takes more time than positron annihilation, the IBD signal forms a coincident pair of prompt and delay. This helps to distinguish the signal from the background. The kinetic energy of the neutron is small, so the initial $\Bar\nu_e$ energy can be roughly calculated by $E_{\Bar\nu_e} \approx E_{\rm prompt} + \SI{0.784}{MeV}$~\cite{dayabaycalib}, where $E_{\rm prompt}$ is the sum of positron kinetic energy and annihilation energy. If $E_{\rm prompt}$ spectrum can be measured precisely, $E_{\Bar\nu_e}$ spectrum can be figured out precisely.

In liquid scintillator detectors, charged particles interact with the liquid scintillator and release photons. However, the number of detected photons is not proportional to the kinetic energy of the charged particle. This effect is referred to as "physics non-linearity" and is mainly caused by ionization quenching and Cherenkov radiation~\cite{dayabaycalib}. In addition, the number of detected photons depends on the position of the charged particle. This is caused by the attenuation of the photons in the liquid scintillator, the solid angle of the photon detector, and the reflection from the material surface, etc~\cite{JUNOcalib}. The position dependence of the number of detected photons is called detector non-uniformity and is generally independent of the particle energy. Non-linearity and non-uniformity are not conducive to the measurement of the precise antineutrino energy spectrum, so both effects need to be corrected by calibration.

The Taishan Antineutrino Observatory (JUNO-TAO, or TAO), which use LS technique, is a satellite detector for JUNO. TAO is located about 30~m from a reactor core of Taishan Nuclear Power Plant. TAO can independently measure the reactor's antineutrino spectrum with unprecedented energy resolution, thereby providing a reference antineutrino spectrum for JUNO. These goals require less than 1\% uncertainty in physics non-linearity and less than 0.5\% residual non-uniformity after correction~\cite{tao_CDR}. In this paper, we mainly present such a calibration strategy that meets these requirements.

The organization of this paper is as follows. In Section~\ref{sec:the_calibration_sys_of_tao}, we introduce the TAO and present the conceptual design of calibration system. In Section~\ref{sec:non-linearity calibration}, we discuss the approach to correct the physics non-linearity. In Section~\ref{sec:nonuniformity_calibration}, we develop a method to correct the non-uniformity and minimize the residual non-uniformity in TAO. Finally, we give a conclusion in Section~\ref{sec:conclusion}.

\section{The calibration system of the TAO experiment}
\label{sec:the_calibration_sys_of_tao}
\subsection{The TAO experiment}
\label{sec:tao_experiment}
\begin{figure}[t]
\centering
\includegraphics[width=0.48\textwidth]{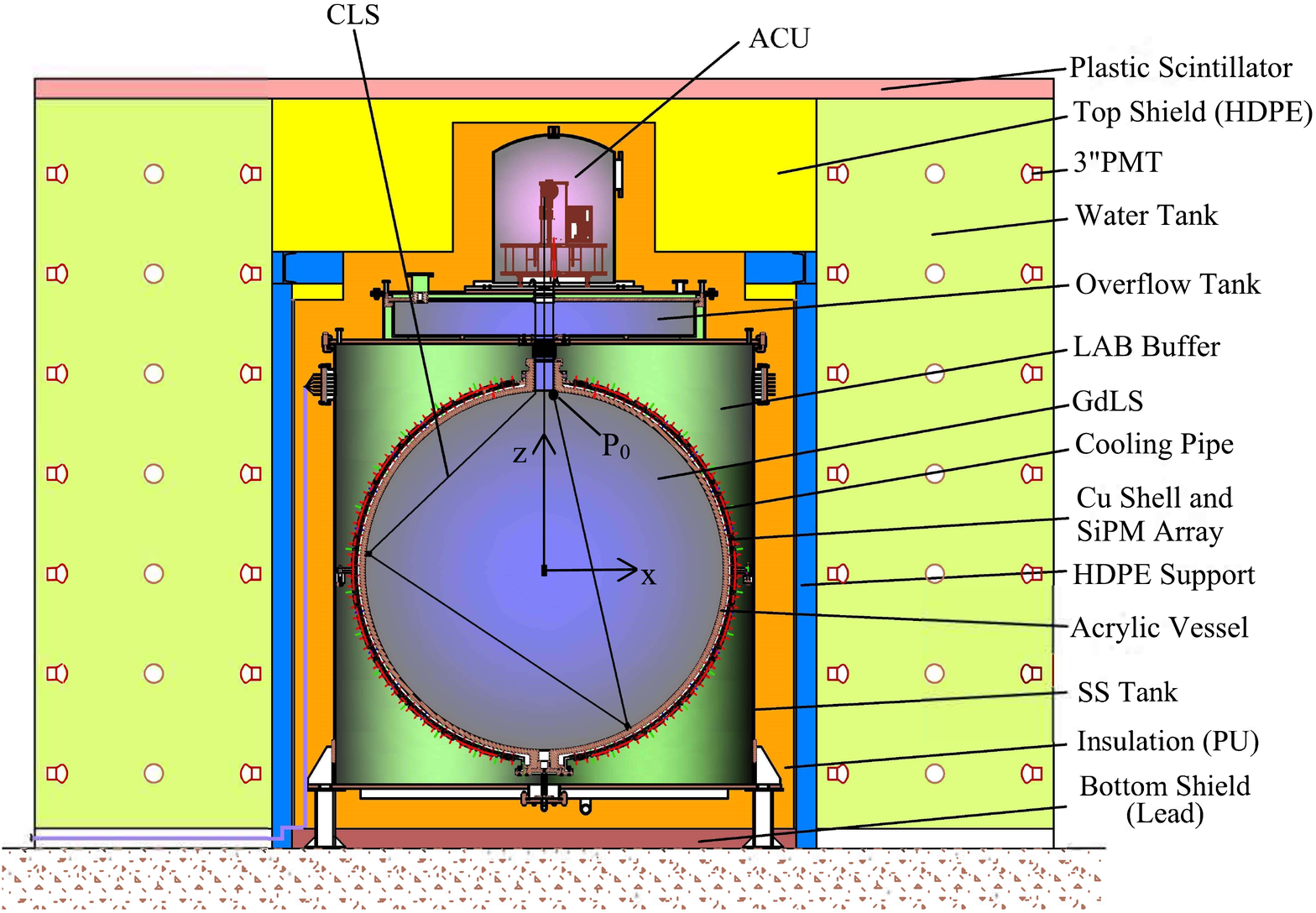}
\caption{Schematic of the TAO detector, which consists of central detector, calibration system, outer shielding and veto system. The Calibration system consists of the Automated Calibration Unit (ACU) and a Cable Loop System (CLS). A segment of the CLS cable is located in GdLS and the start of the segment is marked $\rm P_{\rm 0}$. A coordinate system is specified for the detector. Its origin is at the center of the central detector and its Z-axis points upwards.}
\label{fig:detector}
\end{figure}

As shown in Figure~\ref{fig:detector}, TAO consists of Central Detector (CD), calibration system, outer shielding and veto system. In the CD, there is a spherical acrylic vessel with an inner diameter of \SI{1.8}{m}, which contains approximately \SI{2.8}{tons} of Gadolinium-doped Liquid Scintillator (GdLS). The acrylic vessel is covered by about \SI{10}{\metre\squared} Silicon Photo-multipliers (SiPMs) with high photon detection efficiency. To reduce the dark noise of SiPMs, the detector will operate at \SI{-50}{\degreeCelsius}. To fully contain the energy deposition of gammas from IBD positron annihilation, events within \SI{25}{cm} of the edge of the detector are excluded in the IBD selection, resulting in \SI{1}{ton} fiducial mass. Full detector simulation shows about 4500 photo-electrons per MeV can be detected by SiPMs, leading to excellent energy resolution.


\subsection{Calibration system}
\label{sec:calib_system}
The calibration system contains the Automated Calibration Unit~(ACU) which is reused and modified from the Daya Bay experiment~\cite{dayabay2014acu} and a Cable Loop System~(CLS), as shown in Figure~\ref{fig:detector}. The ACU and CLS can calibrate the energy response on and off the central axis~(Z-axis), respectively.

\begin{figure}[htbp!]
\centering
\subfloat[]{
\includegraphics[width=0.48\textwidth]{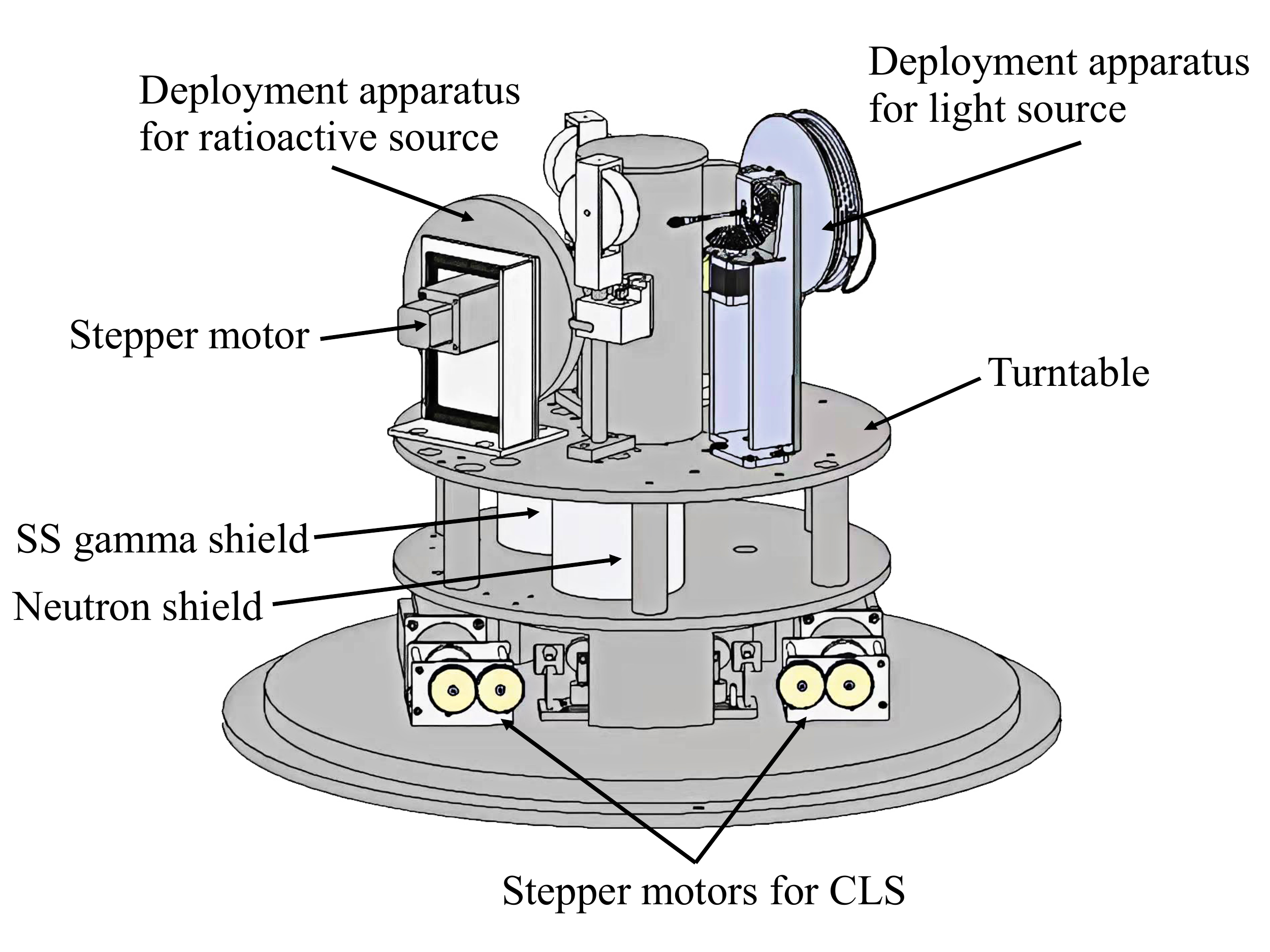}
\label{fig:acu}
}
\hfill
\subfloat[]{
\includegraphics[width=0.28\textwidth]{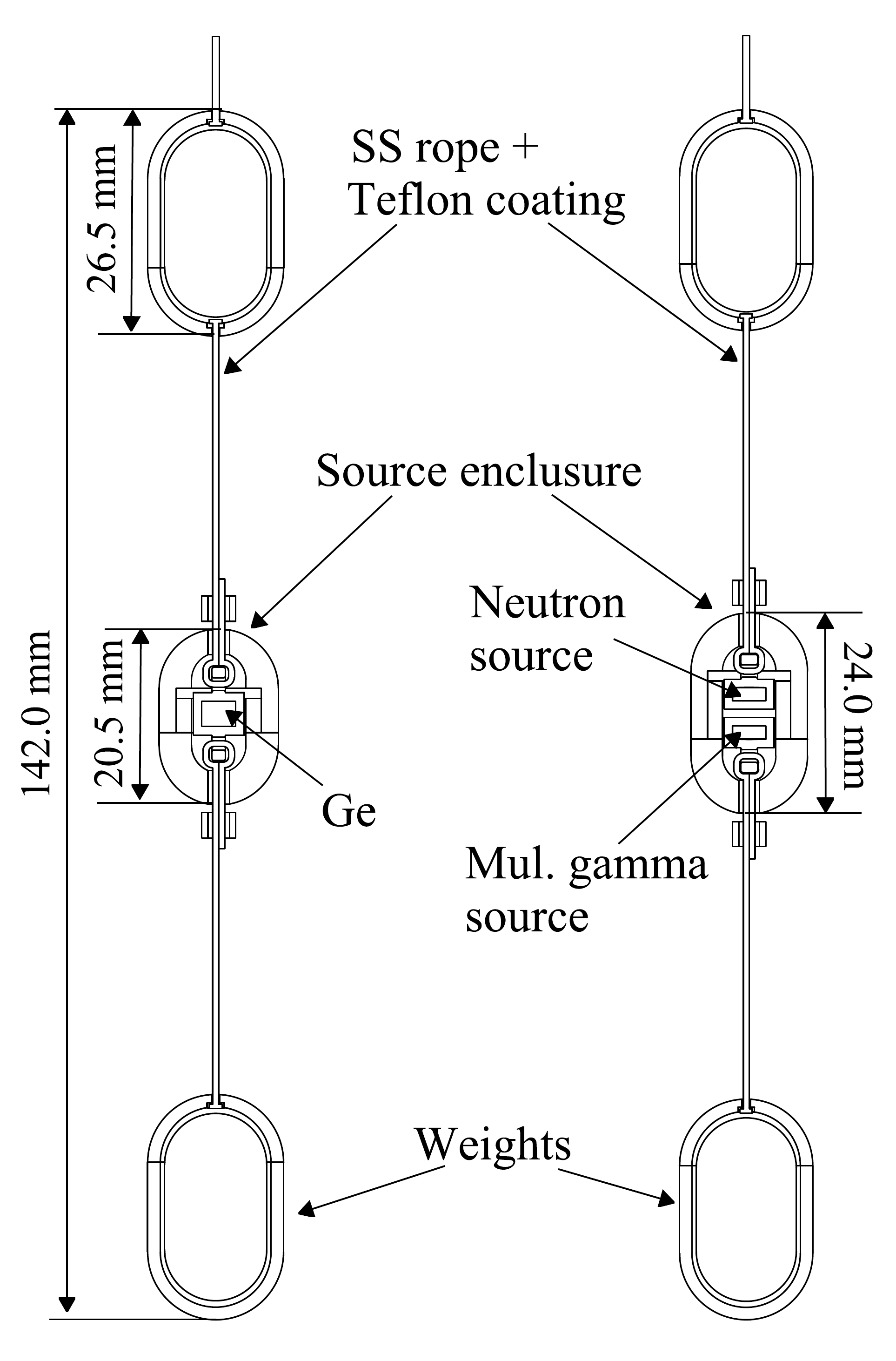}
\label{fig:source_enclosure}
}
\caption{(a) Automated calibration system. (b) Design of $^{68}$Ge source (left) and combined radioactive source (right). The combined source contains multiple gamma sources ($^{137}$Cs, $^{54}$Mn, $^{40}$K, $^{60}$Co) and one neutron source ($^{241}$Am-$^{13}$C).}
\label{fig:acu_total}
\end{figure}

As shown in Figure~\ref{fig:acu}, the ACU consists of a turntable and three mechanically independent motor/pulley/wheel assemblies. The turntable comprises two plates (known as top and middle plates), and three assemblies that are mounted on the top plate. Each assembly is capable of deploying a source into the detector along the central Z-axis once the turntable revolves to a specific angle,  similar to the applications in Daya Bay and JUNO~\cite{dayabay2014acu,juno2021acu}. An ultraviolet (UV) light source, a $^{68}$Ge source, and a combined source that contains multiple gamma sources and one neutron source will be installed on three assemblies, one source for each assembly. 

The $^{68}\rm Ge$ source and the combined source consist of radioactive materials, stainless steel enclosure and Teflon coating as shown in Figure~\ref{fig:source_enclosure}. The Teflon coating can reduce the absorption of light by the source with high reflectivity of 95\%~\cite{neves2017measurement}. In addition, the Teflon is compatible with the GdLS while the stainless steel is not. Besides, each source has one or two counterweights, which allows the wire to maintain tension when the source is deployed to liquid scintillator. When $^{68}$Ge and the combined source are parked in the ACU, they are placed within a stainless steel shield and Borated Polyethylene (BPE) shield respectively as shown in Figure~\ref{fig:acu}. These two shields are cylinders with a wall thickness of \SI{2.25}{inch} and a height of \SI{5}{inch}. Below the BPE shield, there is another BPE disk with a diameter of \SI{3.25}{inch} and a height of \SI{6.4}{inch}. These shields decrease the probability of radioactive products going into the GdLS to a negligible level compared to other backgrounds in the TAO detector.

The UV light source is equipped with a special diffuser that improves the isotropy of its radiation. The wavelength of the UV light source is 265 nm by default and can be changed to 420 nm or any other values if needed. The UV light source can be used to monitor the stability of the parameters of the TAO detector. This task includes monitoring the state of each channel and calibrating its timing, SiPM gain, and quantum efficiency. The UV light source can also be used to test the data acquisition and offline analysis pipeline and to study the pileup in CD.

The CLS allows us to calibrate the detector response in the off-central axis region as showed in Figure~\ref{fig:detector}. 
It adopts some experience from JUNO CLS~\cite{JUNOcalib,zhang2021CLS}. 
The CLS includes two stepper motors, two anchors, a stainless steel cable, load cells, and limit switches. All equipment can work at low temperatures down to \SI{-60}{\degreeCelsius}. The radioactive source ($^{137}$Cs) is plated to a small area of the stainless steel cable, which is coated with Teflon along its entire length to prevent contamination of the GdLS. The anchors made by acrylic are glued to the inner surface of the acrylic vessel in the central detector. The cable is passed through the anchors and can be pulled in either direction by two stepper motors to send the radioactive source into the detector with good positional accuracy. The positions of the anchors are optimized so that we can use the limited calibration positions along the cable to obtain comprehensive information about the non-uniformity response of the detector. When the calibration is complete, the radioactive source is pulled inside the ACU. The limit switch is used to move the radioactive source to the zero position. The load cell is used to monitor the tension of the stainless steel cable and avoid abnormal situations with excessive tension.

\subsection{Simulation software}
The software for simulating TAO is based on SNiPER~\cite{sniper} and Geant4 (10.04.p02)~\cite{geant4}. It contains the geometry of the TAO and the parameters of the physical processes. The optical parameters of GdLS such as the absorption and re-emission probability are taken from Daya Bay software~\cite{dayabay2012sidebyside} since TAO and Daya Bay use similar GdLS and these optical parameters remain unchanged at \SI{-50}{\degreeCelsius} compared to that at room temperature~\cite{xie2021liquid}. The light yield of liquid scintillation increases at low temperatures~\cite{xie2021liquid}, this parameter is adjusted in the simulation. The "Livermore Low Energy" model is used to discribe the electromagnetic physical processes for photons, electrons, hadrons and ions~\cite{geant4physics}. Electronics effects are not yet included in the simulation. This software is used to perform calibration simulation.

\section{Non-linearity calibration}
\label{sec:non-linearity calibration}
In this section, we first introduce the non-linearity model. Then we describe the artificial radioactive sources and natural radioactivity which are used to calibrate physics non-linearity. Next, the systematic biases and uncertainties of the visible energy of these calibration sources are discussed one by one. Finally, we apply the model to fit the calibration data and obtain the calibration performance.

\subsection{Model of physics non-linearity}
Based on the number of photo-electrons ($N_{\rm PE}$) detected by SiPMs, visible energy is defined as
\begin{equation}
    E_{\rm vis} = N_{\rm PE}/Y_{\rm 0},
\label{eq:center_reconstruct}
\end{equation}
where $Y_{\rm 0}$ is the photo-electron yield, equal to \SI{4445}{PE/MeV}. It is determined by simulating the capture of neutrons on hydrogen nuclei at the CD center, and dividing the average of the detected PEs by the gamma energy, namely \SI{2.22}{MeV}. The visible energy of the prompt event of the IBD reaction can be decomposed as
\begin{equation}
    E_{\rm vis}^{\rm prompt} = E^{\rm e}_{\rm vis} + E^{\rm anni}_{\rm vis}.
\end{equation}
$E^{\rm e}_{\rm vis}$ is the visible energy associated with the positron kinetic energy, and it is approximately equal to the electron visible energy with the same kinetic energy \cite{dayabaycalib}. $E^{\rm anni}_{\rm vis}$ is the visible energy of annihilation gammas which can be calibrated by the $^{68}$Ge source. The physics non-linearity of electron or positron is defined by
\begin{equation}
\label{eq:nonlin_definition}
    f_{\rm nonlin}^{\rm e} = E^{\rm e}_{\rm vis}/E^{\rm e},
\end{equation}
where $E^{e}$ is the true kinetic energy of electron or positron. 

\begin{figure}[htbp!]
\centering
\subfloat[]{
\includegraphics[width=0.48\textwidth]{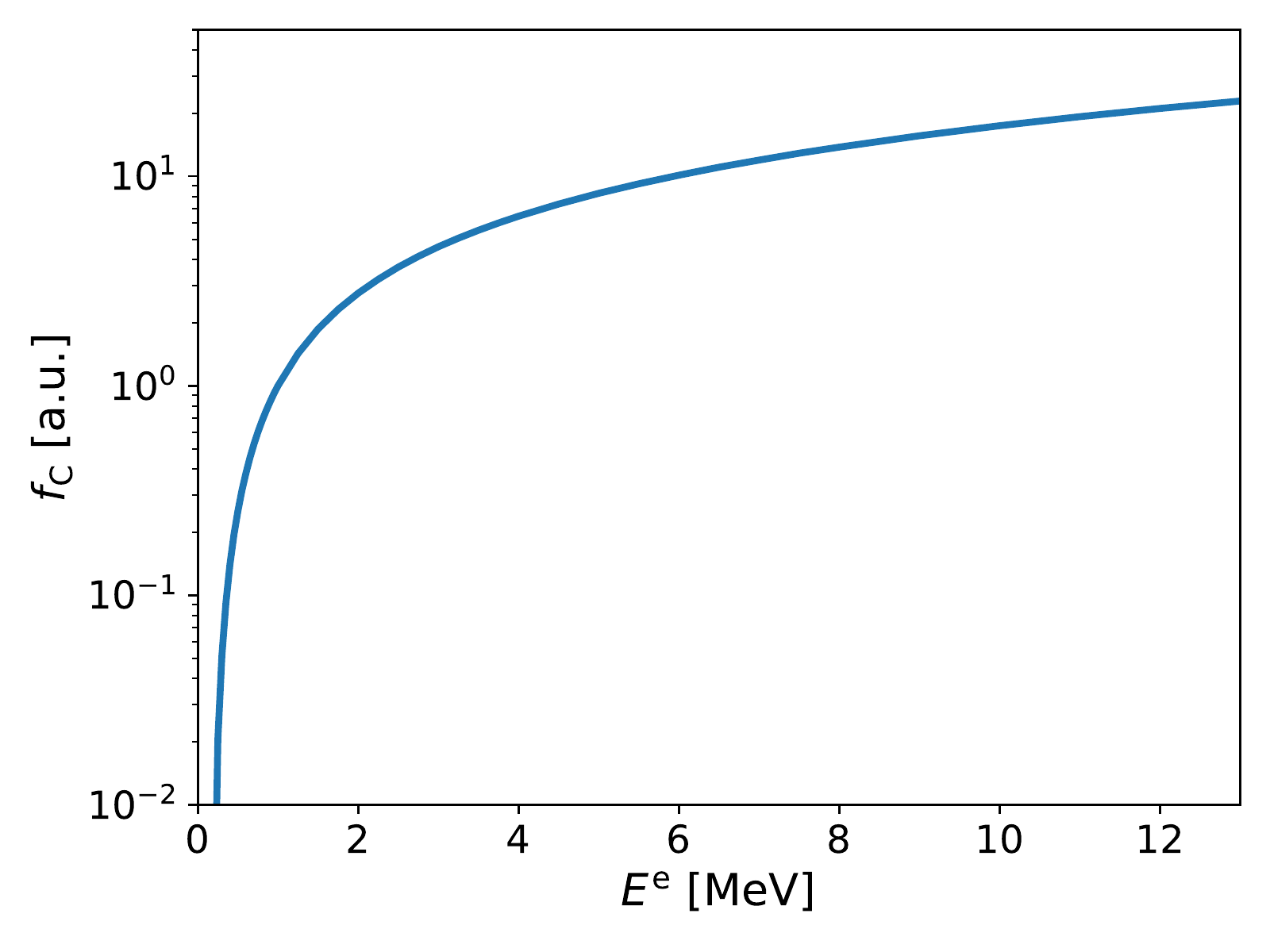}
\label{fig:cherenkov_template}
}
\hfill
\subfloat[]{
\includegraphics[width=0.48\textwidth]{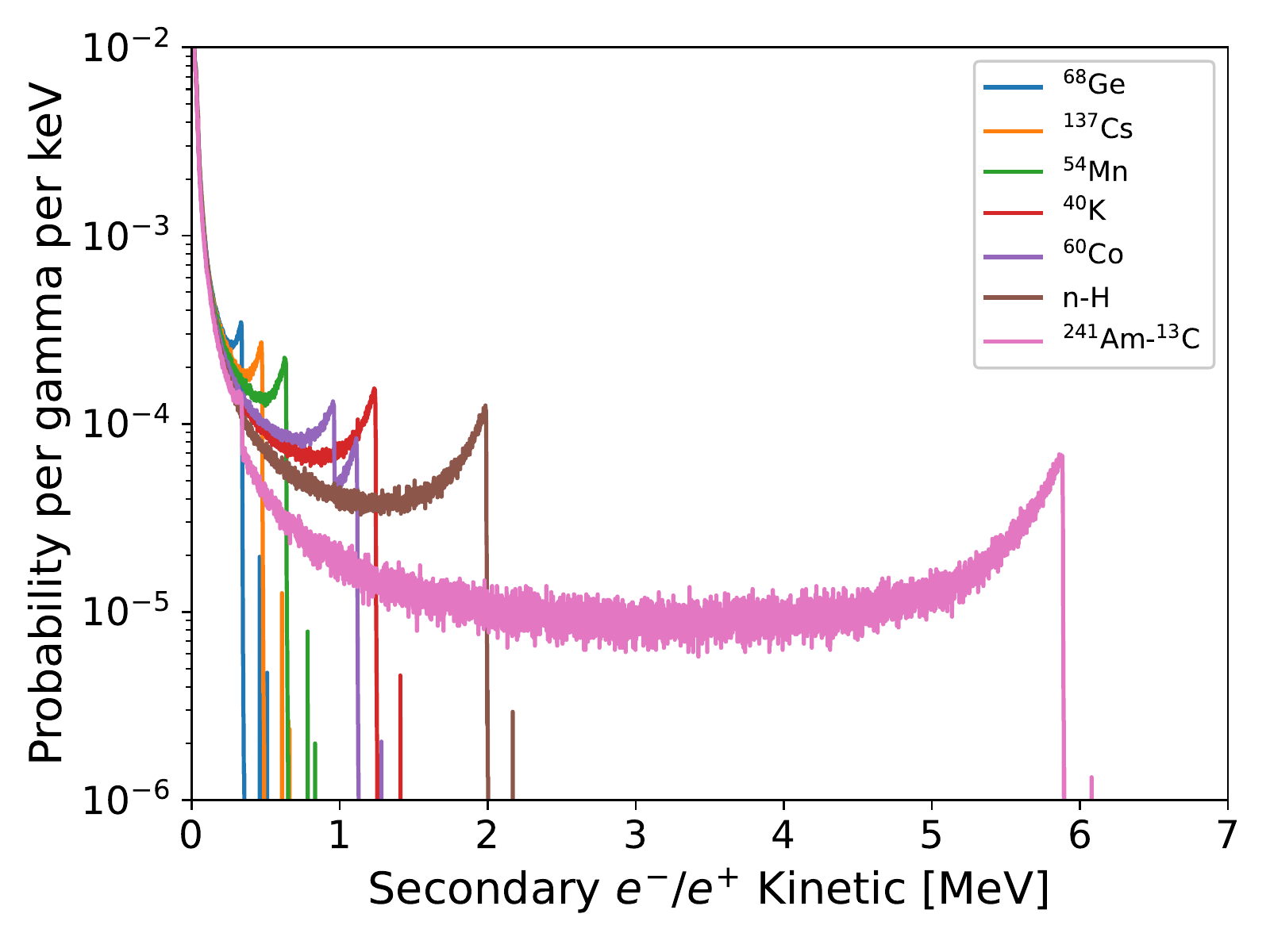}
\label{fig:gamma_electron_distribution}
}
\caption{(a) Number of photo-electrons contributed by Cherenkov radiation. $f_{\rm C}$ is set to 1 at \SI{1}{MeV} and the absolute contribution of Cherenkov radiation can be determined by calibration data. (b) The kinetic energy distributions of secondary electron or positron for a variety of gammas.}
\end{figure}

The physics non-linearity is caused by ionization quenching and the emission of Cherenkov radiation.
\paragraph{Ionization quenching} When the particles deposit energy in GdLS, the solvent molecules are excited, and then the energy is transferred to the fluorescent molecules through dipole-dipole interactions \cite{dayabaycalib}. But when the density of ionized and excited molecules is high, some energy is not transferred. This is quenching effect and causes the non-linear relationship between energy converted to scintillation photons ($E_{\rm scint}$) and deposited energy ($E_{\rm dep}$) of ionizing particle. This quenching effect can be described by Birks' empirical formula~\cite{birks2013theory}:
\begin{equation}
    E_{\rm scint}(E_{\rm dep},k_{\rm B}) = \int\limits_{0}^{E_{\rm dep}}\frac{dE}{1+k_{\rm B}\cdot \frac{dE}{dx}},
\end{equation}
where $k_{\rm B}$ is the Birks' coefficient, and $dE/dx$ is the stopping power. $dE/dx$ is obtained from an ESTAR calculation \cite{berger1999estar} using TAO liquid scintillator properties. 
\paragraph{Cherenkov radiation} Cherenkov photons are produced if the phase velocity of light in the medium is less than the velocity of a charged particle~\cite{Cherenkov:1934}. Since the wavelength spectrum of Cherenkov photons has little dependence on energies of the primary particles, we assume that the number of photo-electrons contributed by Cherenkov radiation is a function of $E^{e}$. The function, referred to as $f_{\rm C}$, is shown in Figure~\ref{fig:cherenkov_template}, which is generated by the Geant4 simulation. $f_{\rm C}$ is set to 1 at \SI{1}{MeV} and the absolute contribution of Cherenkov radiation can be determined by calibration data.

Totally, 
\begin{equation}
    f_{\rm nonlin}^{\rm e}(E^{\rm e};A,k_{\rm B},k_{\rm C}) = A\cdot \left(f_{\rm q}(E^{\rm e},k_{\rm B}) + k_{\rm C}\cdot \frac{f_{\rm C}(E^{\rm e})}{E^{\rm e}}\right),
\label{eq:electron_nonlin}
\end{equation}
where $f_{\rm q}$ is the quenching curve with a Birks' coefficient $k_{\rm B}$, $k_{\rm C}$ is normalization factor of the Cherenkov contribution $f_{\rm C}$, $A$ accounts for the absolute energy scale.

A gamma particle deposits its energy into LS via secondary electrons, so that physics non-linearity of the energy scale in case of gamma radiation can be written as 
\begin{equation}
\begin{aligned}
    f_{\rm nonlin}^{\gamma}&(E^{\rm e};A,k_{\rm B},k_{\rm C}) \equiv \frac{E_{\rm vis}^{\gamma}}{E^{\gamma}} \\
    &= \frac{\int\limits_0^{E_{\gamma}}P(E^{\rm e})\cdot f_{\rm nonlin}^{\rm e}(E^{\rm e};A,k_{\rm B},k_{\rm C})\cdot E^{\rm e}\cdot dE^{\rm e}}{\int\limits_0^{E_{\gamma}}P(E^{\rm e})\cdot E^{\rm e}\cdot dE^{\rm e}},
\end{aligned}
\label{eq:electron_nonlin_model}
\end{equation}
where $P(E^{\rm e})$ denotes the probability density function of a given gamma converted to secondary electron or/and positron with kinetic $E^{\rm e}$ via Compton scattering, photoelectric effect, or pair production \cite{JUNOcalib}. $P(E^{\rm e})$ is determined from simulation, as shown in Figure~\ref{fig:gamma_electron_distribution}.

As for continuous spectrum like $^{12}$B $\beta$-decay spectrum, we can also calculate its expected visible energy distribution as
\begin{equation}
    P_{\rm v}(E_{\rm vis}) = \left( P_{\rm k}(E^{\rm e}(E_{\rm vis})) \cdot \left| \frac{dE^{\rm e}}{dE_{\rm vis}} \right| \right) \otimes {\rm Res}(E_{\rm vis}),
\label{eq:continuous_evis}
\end{equation}
where $P_{\rm v}(E_{\rm vis})$ and $P_{\rm k}(E^{\rm e})$ mean the visible energy and kinetic energy distributions of continuous $\beta$ spectrum respectively, and Res$(E^{\rm e})$ is the energy resolution of the detector. $E^{\rm e}(E_{\rm vis})$ can be calculated by Equation~\ref{eq:nonlin_definition}.

This physics non-linearity model is similar to the model used in the energy scale calibration of Daya Bay experiment \cite{dayabaycalib} which uses similar GdLS.

\subsection{Selection of sources}
\label{sec:select_sources}
The radioactive sources and processes considered in TAO are listed in Table~\ref{tb:radioactive_sources}. We put as many radioactive sources as possible into the ACU, and the gamma energy ranges from a few hundred keV to a few MeV to cover the energy range of the IBD prompt events.

\begin{table}[h!]
\centering
\caption{List of radioactive sources to be used in the TAO non-linearity calibration.}
\label{tb:radioactive_sources}
\scalebox{0.87}{
\begin{tabular}{c|c|c|c}
\hline
Source                     & Type       & Radiation                          & Activity {[}Bq{]} \\ \hline
$^{137}$Cs                 & $\gamma$   & \SI{0.662}{MeV}                               & 50                \\ \hline
$^{54}$Mn                  & $\gamma$   & \SI{0.835}{MeV}                               & 50                \\ \hline
$^{60}$Co                  & $\gamma$   & $\SI{1.173}{MeV} + \SI{1.333}{MeV}$                     & 10                \\ \hline
$^{40}$K                   & $\gamma$   & \SI{1.461}{MeV}                               & 10                \\ \hline
$^{68}$Ge                  & $e^{+}$    & annihilation $\SI{0.511}{MeV} + \SI{0.511}{MeV}$        & 500               \\ \hline
$^{241}$Am-$^{13}$C    & $n,\gamma$ & neutron + \SI{6.13}{MeV} ($^{16}{\rm O}^{*}$) & 2 (neutron)        \\ \hline
$n$($p,\gamma$)$d$             & $\gamma$   & \SI{2.22}{MeV}                                & 2 (neutron)        \\ \hline
\end{tabular}
}
\end{table}

As discussed in Section~\ref{sec:calib_system}, the ACU contains only three independent deployment assemblies for inserting radioactive sources into the detector. We plan to put $^{137}$Cs ,$^{54}$Mn, $^{40}$K, $^{60}$Co, and $^{241}$Am-$^{13}$C into one source enclosure called the "combined source" and put $^{68}$Ge into another source enclosure, remaining one deployment assembly for UV light source. The reason $^{68}$Ge is left alone is that it has a half-life of only 271 days and needs to be replaced after three years. The natural abundance of $^{40}$K is only about 0.012\%~\cite{k40decay} so that enriched $^{40}$K is used to reduce the volume and mass of the radioactive source. The Activities settings of $^{137}$Cs ,$^{54}$Mn, $^{40}$K, and $^{60}$Co ensure that the fully absorbed energy peak of each gamma is not affected much by other gammas as shown in Figure~\ref{fig:combine_source_fit}. When the combined source parks in the ACU, neutrons from $^{241}$Am-$^{13}$C may pass through the shield and enter the central detector. To control the background signals caused by these neutron to an acceptable level, we set the neutron rate of $^{241}$Am-$^{13}$C to 2~Bq~\cite{liu2015neutron}. We plan to deploy the combined source at the detector center for about 10 hours to accumulate enough statistics for each energy peak. Signals from $^{241}$Am-$^{13}$C can be selected by prompt delayed signal correlation and accidental background can be removed by offset-window method \cite{dayabay2017measurement}. Thus, we can get visible energy spectrum of $^{16}$O$^{*}$ gamma events and neutron capture events.

 The simulation shows that the muon rate in TAO CD is about \SI{330}{\second^{-1}}. The energetic cosmic muons and subsequent showers can interact with $^{12}$C in GdLS to produce unstable isotopes like $^{12}$B~\cite{juno2016neutrinophys}. $^{12}$B decays via $\beta$-emissions with a mean life time of \SI{20.2}{ms} and with a $Q$ value of \SI{13.4}{MeV}, so the $^{12}$B events can provide a constraint on the electron non-linearity model, especially in high energy range. In order to select the $^{12}$B events, we firstly select neutron tagged muons whose rate is smaller than \SI{1}{\second^{-1}} and then apply a time window after each muon~\cite{dayabaycalib}. Approximately $10^5$ $^{12}$B events can be selected in the fiducial volume in three years. About 2.8\% $^{12}$N decay events is mixed into the energy spectrum of $^{12}$B. The change of the fitted physics non-linearity (Equation~\ref{eq:electron_nonlin}) is less than 0.5\textperthousand{} when $^{12}$N events are taken into consideration. For the sake of simplicity, $^{12}$N events are ignored here.

\subsection{Systematic biases and uncertainties}
In this section, we analyze systematic biases of visible energy of calibration sources. The effects that lead to systematic biases are analyzed one by one. These systematic biases are assumed to be correctable but with conservative 100\% uncertainties. This analysis refers to the experience of the Daya Bay \cite{dayabaycalib} and the JUNO \cite{JUNOcalib} experiments.

\subsubsection{Energy loss effect}

\begin{figure}
\centering
\subfloat[]{
\includegraphics[width=0.48\textwidth]{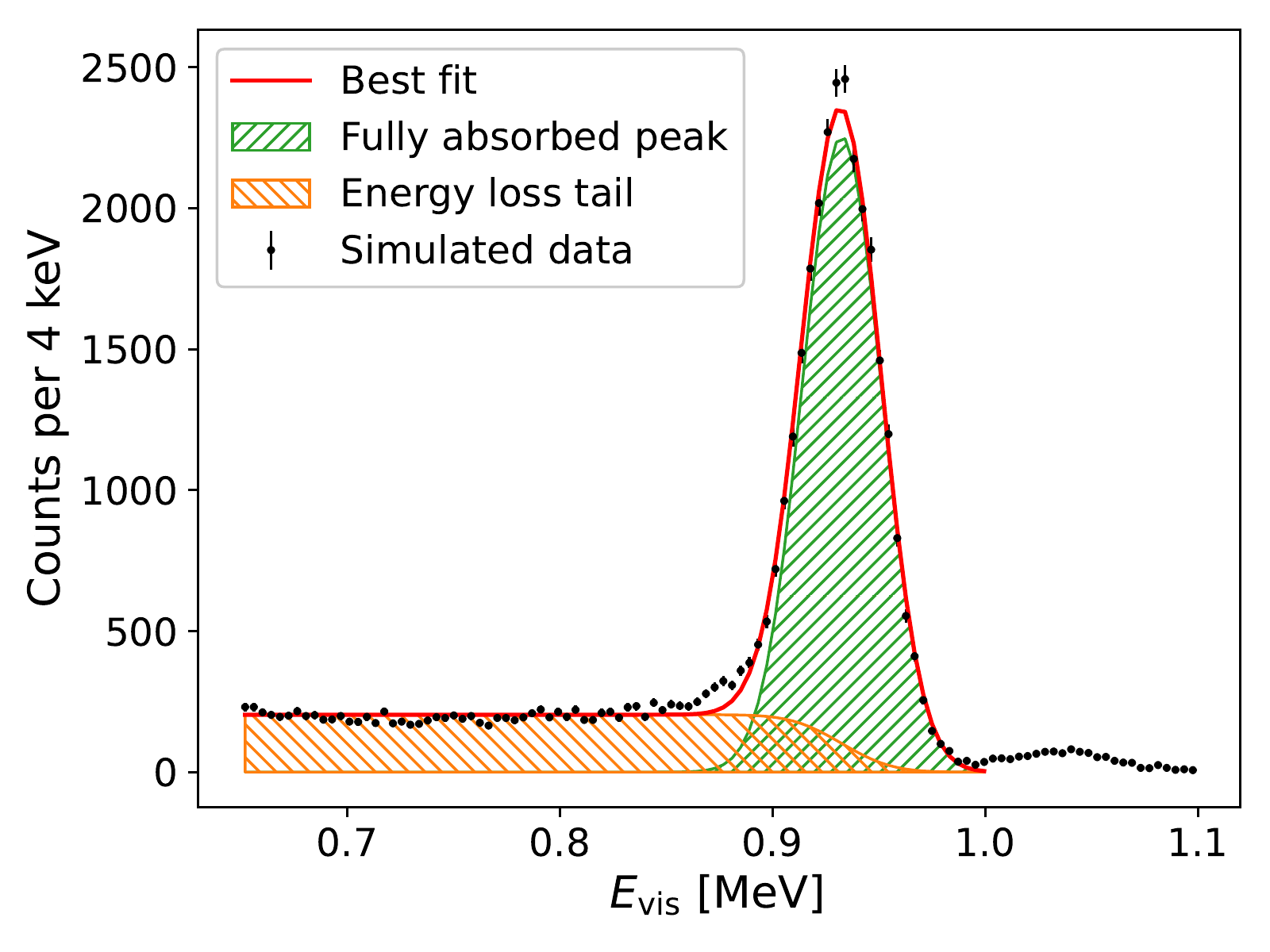}
\label{fig:ge68_fit}
}
\hfill
\subfloat[]{
\includegraphics[width=0.48\textwidth]{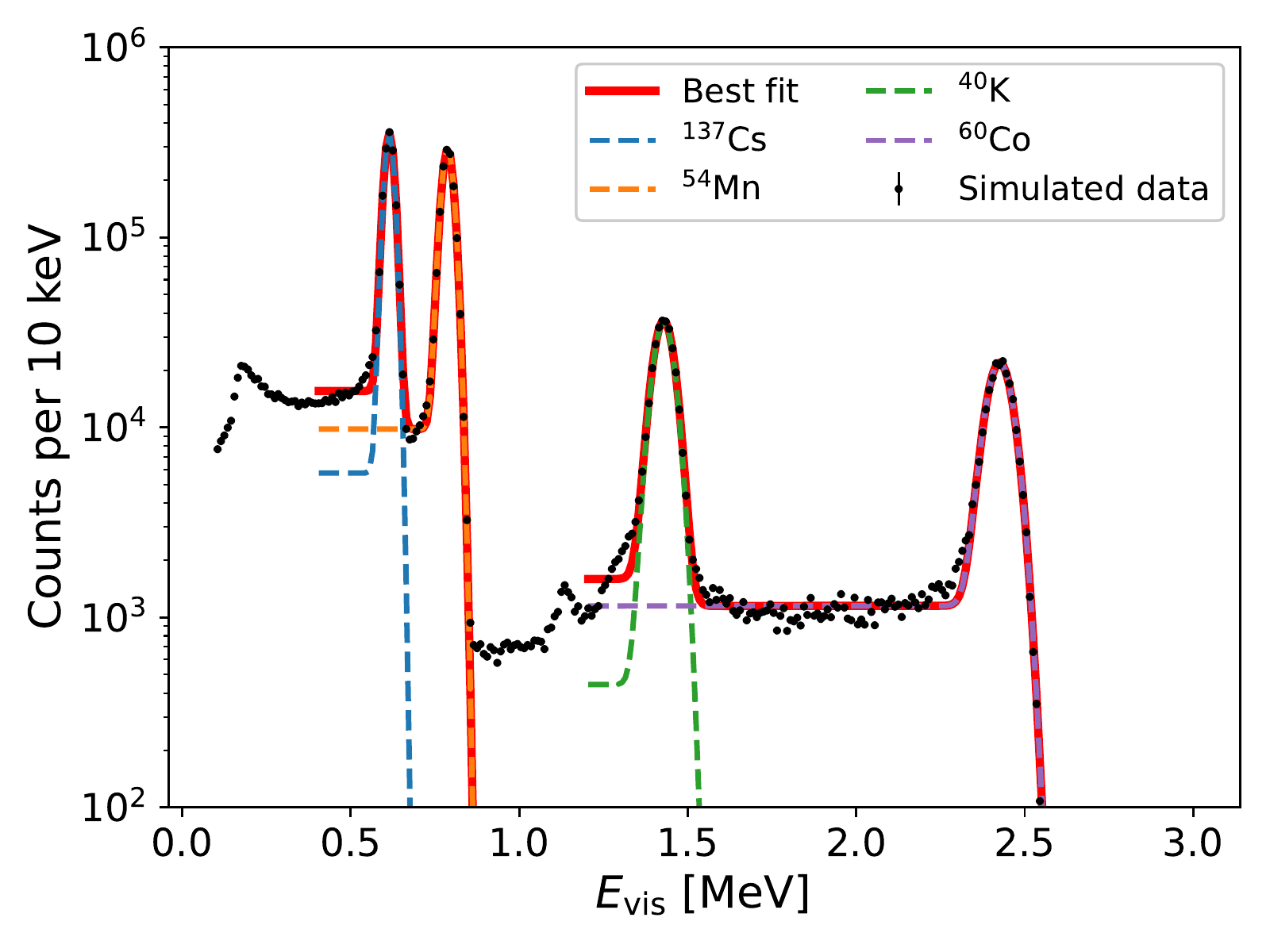}
\label{fig:combine_source_fit}
}
\caption{(a) The visible energy spectra of $^{68}$Ge. The number of $^{68}$Ge events is equivalent to 100 seconds' calibration data. The contributions of the fully absorbed peak and the energy loss tail are shown in the figure. (b) Gamma spectrum and the fitting results of the radioactive isotopes in the combined source. The number of combined source events is equivalent to 10 hours' data. All the fully absorbed peaks are distinguishable. The spectrum of $^{137}$Cs and $^{54}$Mn are fitted separately from the spectrum of $^{40}$K and $^{60}$Co.}
\label{fig:ratioactive_source_fit}
\end{figure}

Gamma particles may deposit some of their energy in non-scintillating materials such as the source enclosure, weights, etc. This results in an energy loss tail in the visible energy distribution, and this tail causes a bias in the fit to the fully absorbed peak. We investigate this energy loss effect by separately simulating the $^{68}$Ge source and the combined source at the center of the CD.

For a single $\gamma$ source, we apply a Gaussian function to model the fully absorbed energy peak and use a complementary error function with a normalization parameter to model the energy loss tail.
\begin{small}
\begin{equation}
    f(E_{\rm vis};\eta_{0},\eta_{1},\mu,\sigma) = \eta_{0}\cdot \left( e^{-\frac{(E_{\rm vis}-\mu)^{2}}{2\sigma^{2}}} + \eta_{1}\cdot {\rm erfc}\left(\frac{E_{\rm vis} - \mu}{\sqrt{2}\sigma}\right)\right),
\label{eq:source_fitting}
\end{equation}
\end{small}

where $\rm erfc$ is the complementary error function. $\mu$ and $\sigma$ represent the mean and standard deviation of visible energy spectrum of fully absorbed peak. The fitted $\mu$ value is marked as $E_{\rm vis}^{\rm fit}$. $\eta_{0}$ is the absolute amplitude of the fully absorbed peak while $\eta_1$ is the relative amplitude of the energy loss tail. As an example, this function is applied to fit the energy spectrum of $^{68}$Ge, as shown in Figure~\ref{fig:ge68_fit}. For comparison, events with energy fully absorbed in the GdLS are selected. A Gaussian function is used to model the visible energy spectrum of these selected events and the fitted mean are marked as $E_{\rm vis}^{\rm ideal}$. We define $(E_{\rm vis}^{\rm fit} - E_{\rm vis}^{\rm ideal})/E_{\rm vis}^{\rm ideal}$ as fitting bias to measure this deviation caused by energy loss. 

For $^{137}$Cs, $^{54}$Mn, $^{40}$K, and $^{60}$Co in the combined source, the total visible energy spectrum is obtained by simulation. As shown in  Figure~\ref{fig:combine_source_fit}, the spectrum of $^{40}$K and $^{60}$Co has little effect on the spectrum of $^{137}$Cs and $^{54}$Mn, so the spectrum of $^{54}$Mn and $^{137}$Cs are fitted separately from the spectrum of $^{40}$K and $^{60}$Co.

As mentioned in Section~\ref{sec:select_sources}, we can also get visible energy spectrum of neutron capture signals. In GdLS, neutron capture mainly happens on Gd nucleus (n-Gd) and H nucleus (n-H). Neutron capture on Gd nucleus will emit multiple gammas and the energy spectrum of these gammas is very model-dependent~\cite{dayabay2018improvefulx}, so only n-H gammas are used for non-linearity calibration. Since energy loss tail of n-Gd events affects the peak of n-H visible energy spectrum, $\eta_{2} + f(E_{\rm vis}; \eta_{0}, \eta_{1}, \mu, \sigma)$ is applied to model the visible energy spectrum near n-H fully absorbed peak, where $\eta_{2}$ models the energy loss tail of n-Gd spectrum. 

\begin{figure}[htbp!]
\centering
\includegraphics[width=0.48\textwidth]{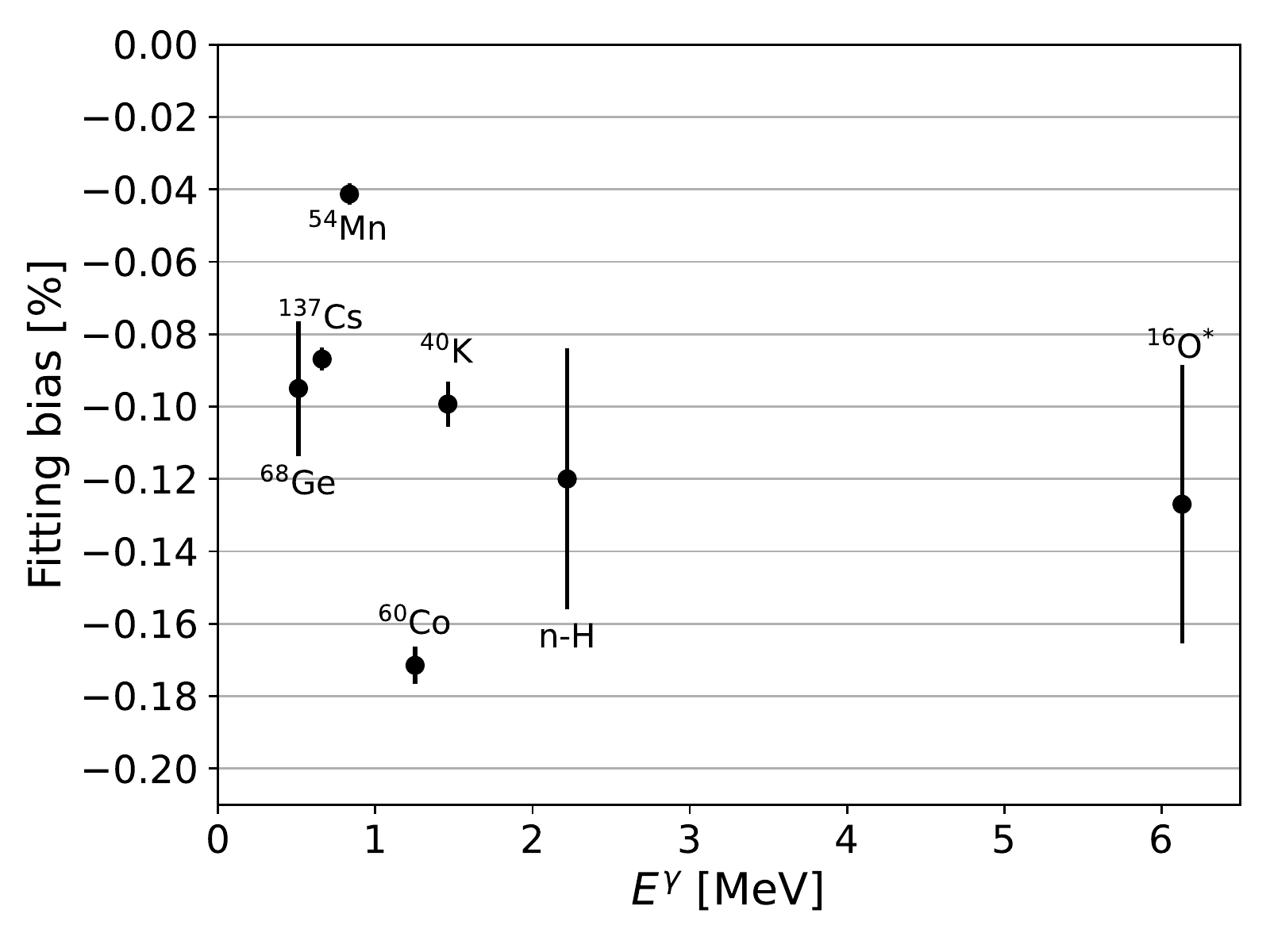}
\caption{The fitting bias due to energy loss effect. It is defined as $(E_{\rm vis}^{\rm fit} - E_{\rm vis}^{\rm ideal})/E_{\rm vis}^{\rm ideal}$.}
\label{fig:fitting_bias}
\end{figure}

As shown in Figure~\ref{fig:fitting_bias}, the fitting bias is less than 0.2\% for radioactive sources. The fitting bias is caused by energy loss. Since the energy losses of different gamma sources are due to similar non-scintillating materials, the fitting biases between these gammas are assumed to be correlated.

\subsubsection{Shadowing effect}
As mentioned before, we need to put radioactive sources into enclosures. Besides, in order to keep the tension of the stainless steel rope when we deploy these radioactive sources into GdLS, two weights are used along with each source enclosure. When UV and optic photons in GdLS meet the surfaces of source enclosures and weights, they can be absorbed. To reduce these effects, the surfaces are covered by Polytetrafluoroethylene (PTFE) whose reflectivity is very high~\cite{neves2017measurement}. In this study, we assume a $95\%$ reflectivity to study the shadowing bias.

\begin{figure}[htbp!]
\centering
\includegraphics[width=0.48\textwidth]{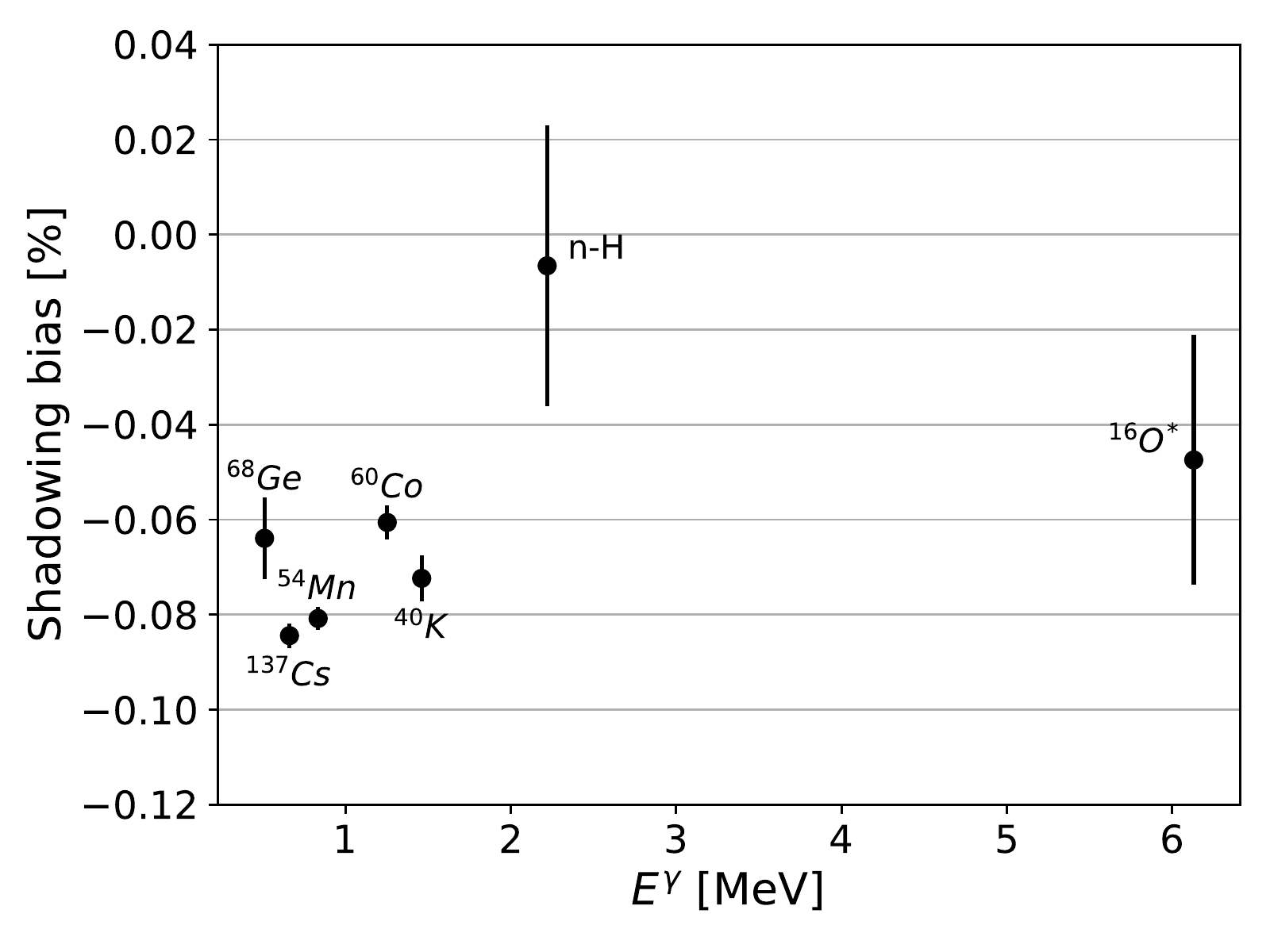}
\caption{The shadowing bias due to UV and optical photons from gammas absorbed by the surfaces of the source enclosure and weights. It is defined as $(E_{\rm vis}^{\rm shadow} - E_{\rm vis}^{\rm nos})/E_{\rm vis}^{\rm nos}$.}
\label{fig:shadowing_bias}
\end{figure}

In order to decouple the energy loss effect, only events with energy fully absorbed by GdLS are used. The visible energy spectrum are modeled by a Gaussian function. The fitted mean of the Gaussian is marked as $E_{\rm vis}^{\rm shadow}$. For comparison, we simulate naked radioactive sources to avoid shadowing effect. The fitted mean of the visible energy spectrum of a naked source is marked as $E_{\rm vis}^{\rm nos}$. Bias caused by shadowing effect is defined as $(E_{\rm vis}^{\rm shadow} - E_{\rm vis}^{\rm nos})/E_{\rm vis}^{\rm nos}$. As shown in Figure~\ref{fig:shadowing_bias}, shadowing bias of each radioactive source is smaller than 0.1\%. It is assumed that shadowing biases are correlated among different radioactive sources because they are caused by the similar enclosures and weights.

\subsubsection{$^{16}\rm O^{*}$ \SI{6.13}{MeV} gamma bias}
When the $^{241}{\rm Am}$-$^{13}$C source releases a gamma with an energy of 6.13~MeV, it also releases a neutron with a kinetic energy of less than \SI{100}{keV}~\cite{dayabaycalib}. This neutron can cause a proton recoil, increasing the number of photo-electrons that are eventually detected by SiPMs. This introduces about a 0.4\% bias to the measurement of the visible energy of 6.13~MeV gamma.

\subsubsection{Residual bias after non-uniformity correction}
Events such as $^{12}$B $\beta$ decays distribute uniformly in the detector and gammas coming from calibration sources do not deposit their energy as point sources. The non-uniformity of the detector affects their visible energy spectrum. The non-uniformity effect can be corrected but not perfectly. According to Section~\ref{sec:energy_resolution}, the residual bias after non-uniformity correction can be controlled within $0.3\%$. It is conservatively taken as a fully correlated bias among different energies.

\subsubsection{Instrumental non-linearity}
The bias caused by non-linearity of the SiPM readout can be neglected. Assuming the size of a single SiPM is 6×6 mm$^2$, the total number of SiPM is about $2.7 \times 10^{5}$ \cite{tao_CDR}. Besides, assuming the size of a SiPM pixel is 60×60 $\mu\rm m^2$, there are $10^{4}$ SiPM pixels on a single SiPM. The simulation shows that, with $\SI{1}{MeV}$ of energy deposited in GdLS, SiPMs collect approximately 4500 photo-electrons. For antineutrino events with an prompt energy range from 1 MeV to 10 MeV, the number of photons hitting a SiPM is much smaller than the number of pixels on the SiPM, so the response of the SiPM is linear. In addition, the charge information can be reconstructed well because the waveform information of the readout channels is stored.

\subsubsection{Summary of systematic bias and uncertainty}
The effects discussed above are summarized in Table~\ref{tb:system_uncertainties}. There are two types of biases, correlated and single point, depending on their impact on visible energies. For example, the biases caused by the shadowing effect between different sources are correlated due to similar source enclosure and weights. On the other hand, the bias of \SI{6.13}{MeV} $\gamma$ is independent from the others (single point). We assume that the biases can be corrected, but with a conservative uncertainty of 100\%~\cite{JUNOcalib}. It is assumed that the type of uncertainty is the same as the corresponding bias.

\begin{table*}[t]
\centering
\caption{A list of systematic biases and uncertainties. The uncertainties are assumed as 100\% of the absolute values of the biases. The last column indicates whether the biases and uncertainties are correlated between different sources or energies.}
\label{tb:system_uncertainties}
\begin{tabular}{c|c|c|c}
\hline
source                       & bias              & uncertainty      & type       \\ \hline
Energy loss effect           & -0.2\%$\sim$-0.04\% & \textless{}0.2\% & correlated   \\ \hline
Shadowing effect             & -0.1\%$\sim$0\%   & \textless{}0.1\% & correlated   \\ \hline
$^{16}\rm O^{*}$ \SI{6.13}{MeV} $\gamma$ uncertainty & +0.4\%            & 0.4\%            & single point \\ \hline
Instrumental non-linearity   & $\sim$0           & $\sim$0          &              \\ \hline
Position-dependent effect    & \textless{}0.3\%  & 0.3\%            & correlated   \\ \hline
\end{tabular}
\end{table*}

\subsection{Non-linearity fitting}
The radioactive sources mentioned in Section~\ref{sec:select_sources} are used to demonstrate the performance of the non-linearity calibration. The combined source and  the $^{68}$Ge source are simulated at the center of the detector. The statistics of combined source and $^{68}$Ge source are calculated with the planned calibration data taking time, \SI{10}{\hour} and \SI{100}{\second} respectively. Besides, $10^5$ $^{12}$B events that can be collected in fiducial volume in three years are simulated. Equation~\ref{eq:reconstruct} is used to reconstruct the visible energy of each event. $^{12}$B events below \SI{3}{MeV}  are ignored due to high background contamination in this range. For simplicity, $^{12}$B events above \SI{12}{MeV} are also ignored because there are many $^{12}$N events in this energy range \cite{dayabaycalib}.

\begin{figure}
\centering
\subfloat[]{
\includegraphics[width=0.48\textwidth]{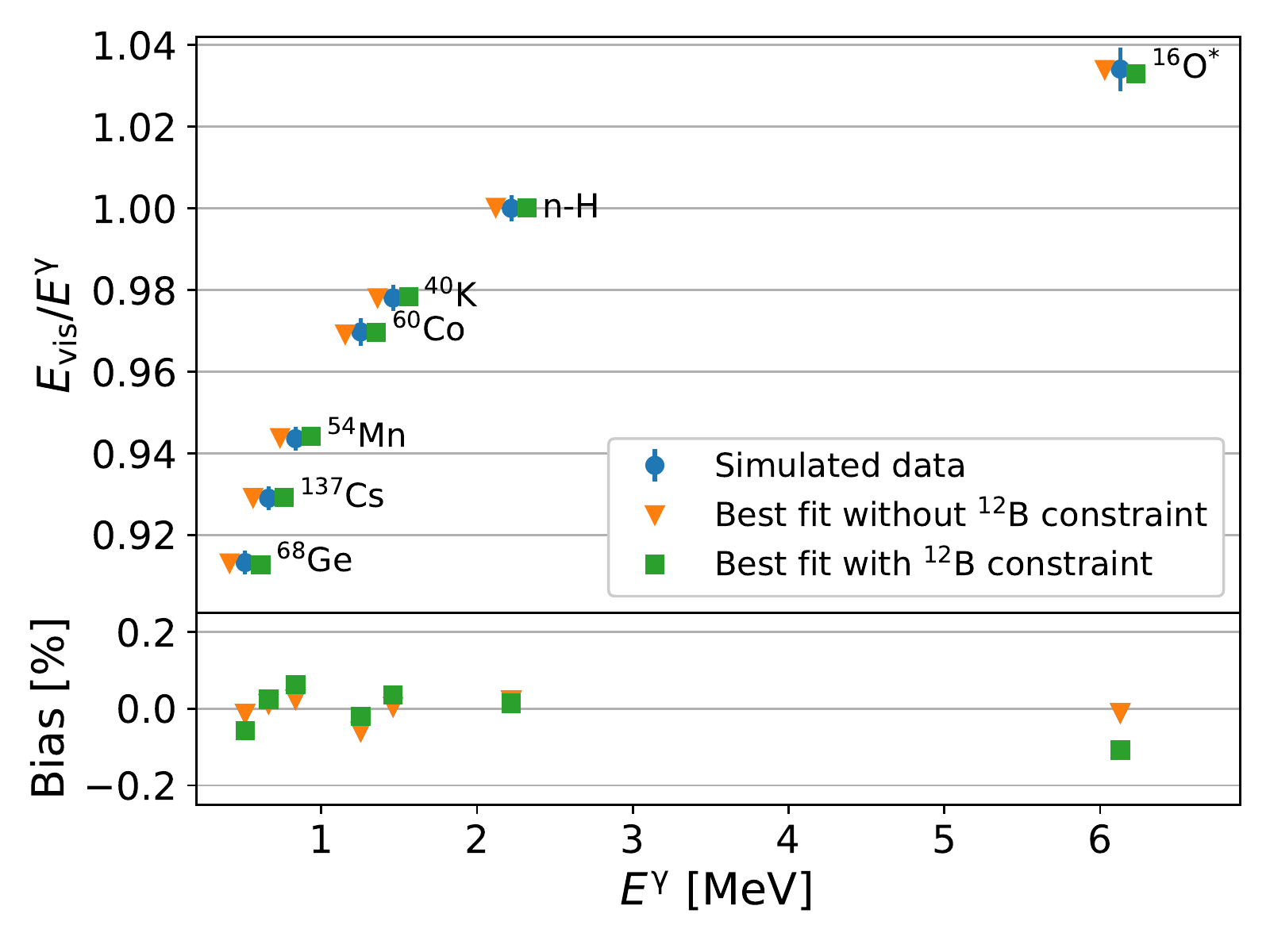}
\label{fig:gamma_fitting}
}
\hfill
\subfloat[]{
\includegraphics[width=0.48\textwidth]{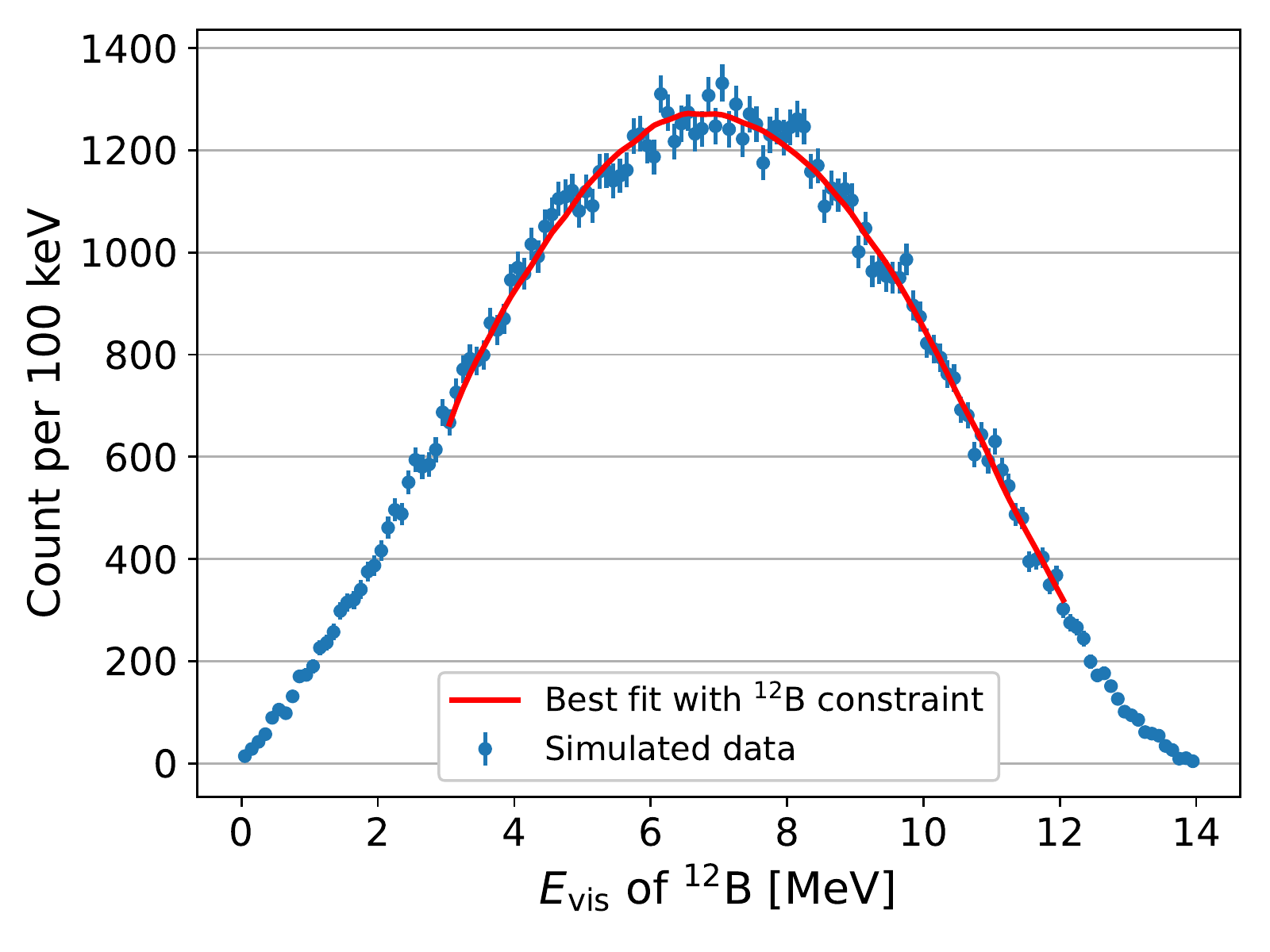}
\label{fig:b12_fitting}
}
\caption{(a) The best fit points of the gamma radioactive sources compared with the simulated data points. There are two fitting cases on the plot, namely with and without $^{12}$B constraint. The fitted data points are shifted on the x-axis to avoid overlapping. (b) Best fit and simulated $^{12}$B spectrum. The total number of $^{12}$B events is about $\rm 10^5$, which can be collected in fiducial volume in 3 years of data taking.}
\label{fig:fitting_performance}
\end{figure}

\begin{figure}
\centering
\includegraphics[width=0.48\textwidth]{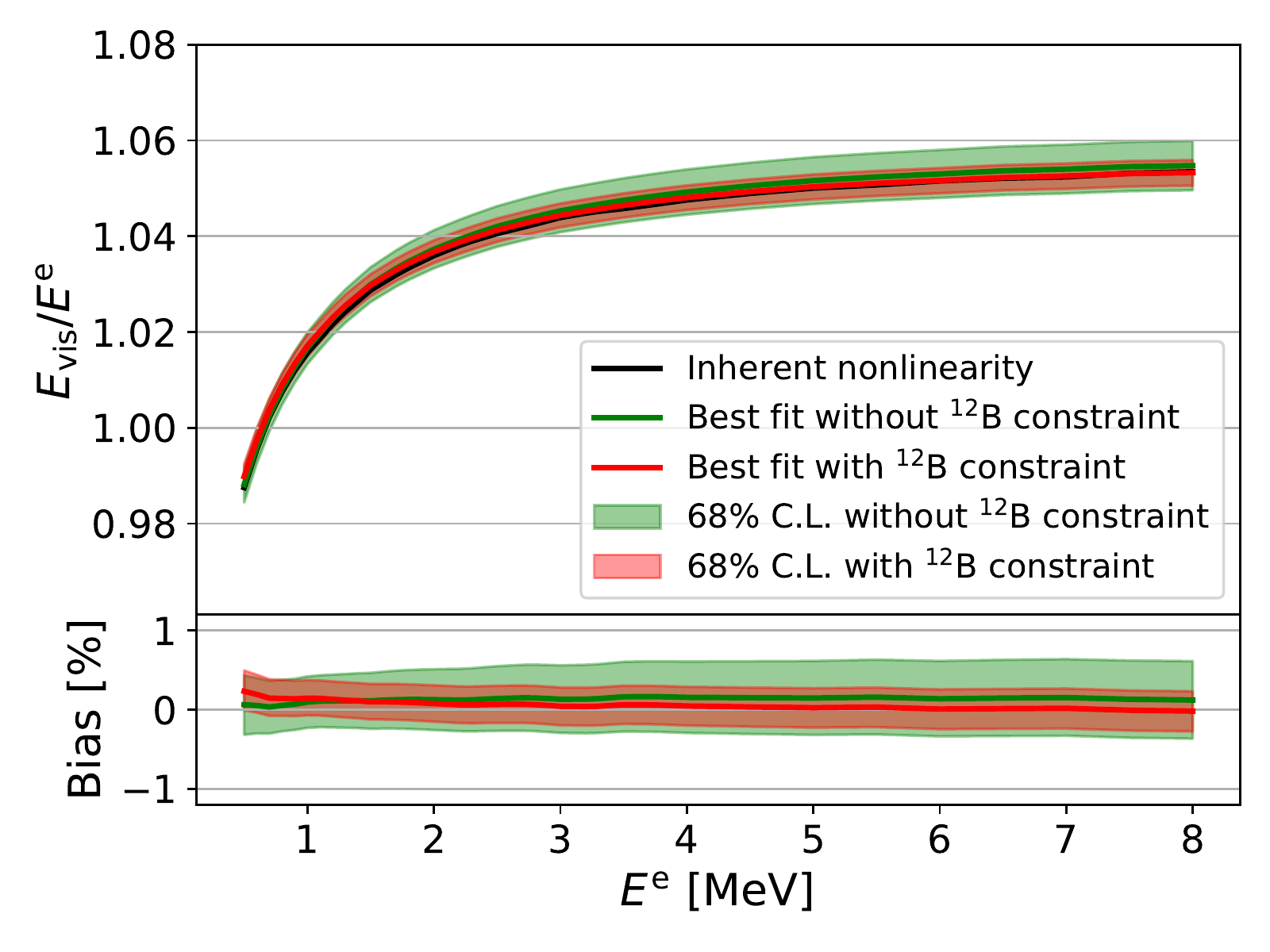}
\caption{Electron non-linearity fitting results. $E^{e}$ is the kinetic energy of electron or positron.  Black line is the inherent non-linearity. Green line and band is the best fit and $68\%$ confidence interval without constraint from $^{12}$B spectrum. Red line and band is the best fit and $68\%$ confidence interval with constraint from $^{12}$B spectrum.}
\label{fig:electron_nonlin}
\end{figure}

To fit the non-linearity model in Equation~\ref{eq:electron_nonlin_model}, $\chi^{2}$ is defined as:
\begin{equation}
    \chi^{2} = \sum_{i=1}^{7}\left(\frac{M_i^{\gamma} - P_i^{\gamma}}{\sigma_i}\right)^{2} + \sum_{j=1}^{90}\frac{\left(M_{j}^{^{12}{\rm B}} - P_{j}^ {^{12}{\rm B}}\right)^{2}}{M_{j}^{^{12}{\rm B}}}.
\label{eq:nonlin_chi2}
\end{equation}
$M_i^{\gamma}$ and $P_i^{\gamma}$ here are measured and predicted visible energy peak of gamma source respectively. $\sigma_i$ contains statistical uncertainty and systematic uncertainties of $M_i^{\gamma}$, where systematic uncertainties are listed in Table~\ref{tb:system_uncertainties}.
In the energy range from 3~MeV to 12~MeV, the energy spectrum of $^{12}$B is equally divided into 90 bins. $M_{j}^{^{12}{\rm B}}$ is the number of $^{12}$B events measured in the $j$-th bin, and $P_{j}^{^{12}{\rm B}}$ is the number of $^{12}$B events calculated by Equation~\ref{eq:continuous_evis} in the $j$-th bin. When there is not enough $^{12}$B events, only the first term in Equation~\ref{eq:nonlin_chi2} is used. Minimizing Equation~\ref{eq:nonlin_chi2}, we obtain best fit values of non-linearity model parameters $A$, $k_{\rm B}$ and $k_{\rm C}$ in Equation~\ref{eq:electron_nonlin_model}. 
The fitting performance is shown in Figure~\ref{fig:fitting_performance}, the difference between the fitted $E^{\gamma}_{\rm vis}$ and the simulated $E^{\gamma}_{\rm vis}$ is less than 0.2\%.

To calculate the 68\% confidence interval, we sample the visible energy of $\gamma$s and $^{12}$B $\beta$ decays according to the $1\sigma$ uncertainties listed in Table~\ref{tb:system_uncertainties} and statistical uncertainty, either in a correlated or single point fashion. For a given correlated uncertainty, a random number that obeys the standard normal distribution is generated and is multiplied by the uncertainty to calculate the offsets for visible energies of calibration sources. For uncorrelated or single point uncertainties, the data points are shifted independently according to 1$\sigma$ uncertainties. We repeat this operation many times and fit each set of data to get many electron non-linearity curves. Finally, we can get 68\% confidence interval for physics non-linearity for each energy as showed in Figure~\ref{fig:electron_nonlin}.

For comparison, mono-energetic electrons are simulated at the CD center to obtain an inherent non-linearity, as shown in Figure~\ref{fig:electron_nonlin}. For the situation without $^{12}$B data constraint, the best fit curve and the inherent non-linearity agrees within 0.2\% in the energy range from \SI{0.5}{MeV} to \SI{8}{MeV}. The uncertainty of the best fit curve is less than 0.6\% in the same energy range. This uncertainty is less than 1\%, which means that our requirement is satisfied. For the situation with three years $^{12}$B data constraint, the uncertainty is less than 0.4\%.

\section{Non-uniformity calibration}
\label{sec:nonuniformity_calibration}
When particles interact with GdLS in the CD at different positions, the detector responses are different. This is the non-uniformity of the detector and can be characterized by $g(r,\theta,\phi)$, which is defined as the photo-electron yield at a given position relative to the photo-electron yield at the center of the detector. r, $\theta$ and $\phi$ are the radius, polar angle and azimuthal angle of the given position in spherical coordinate. The origin of the spherical coordinate is at the center of the CD, and the zenith points upward, as shown in Figure~\ref{fig:detector}. The non-uniformity of the detector degrades the energy resolution, so we need to understand the non-uniformity well.

The key question is how to calibrate $g(r,\theta,\phi)$ well enough in real situations. In this section, we first obtain the non-uniformity of the detector using a perfect electron source that can be inserted into any given position of the detector. Taking this non-uniformity as a reference, we optimize the design of the CLS anchors and the selection of the limited calibration points along the CLS cable. The non-uniformity of the detector is calibrated with gamma sources deployed at these selected calibration points. Once $g(r,\theta,\phi)$ is obtained, the visible IBD prompt energy can be evaluated by
\begin{equation}
    E^{\rm prompt}_{\rm vis}(r,\theta,\phi) = N_{\rm PE}^{\rm tot}/g(r,\theta,\phi)/Y_{\rm 0},
\label{eq:reconstruct}
\end{equation}
where $N_{\rm PE}^{\rm tot}$ is the number of photo-electrons detected by SiPMs and $Y_{\rm 0}$ is the photo-electron yield at the CD center.

\subsection{Ideal non-uniformity map}
\begin{figure}[htbp!]
\centering
\subfloat[]{
\includegraphics[width=0.48\textwidth]{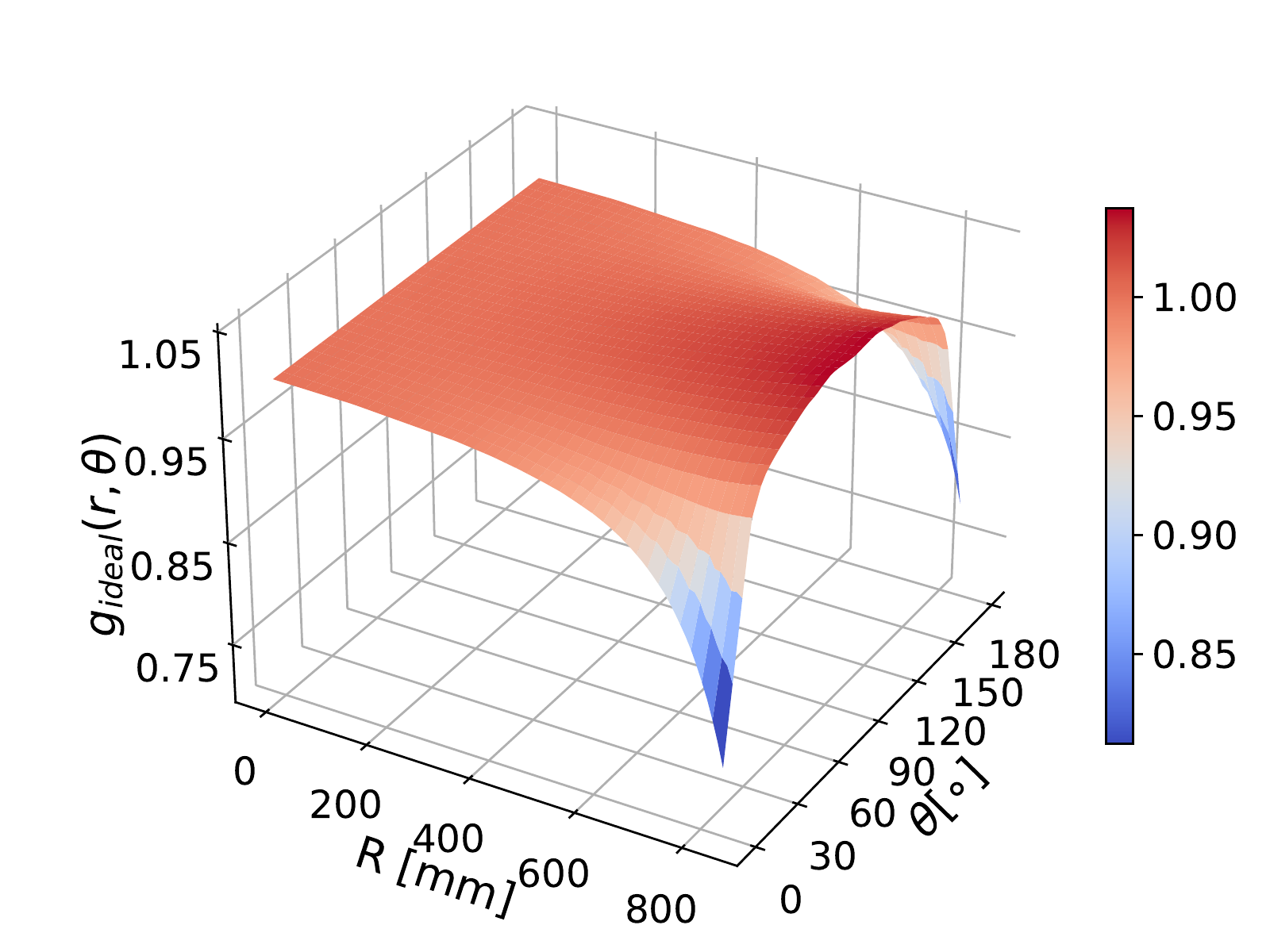}
\label{fig:ideal_numap}
}
\hfill
\subfloat[]{
\includegraphics[width=0.48\textwidth]{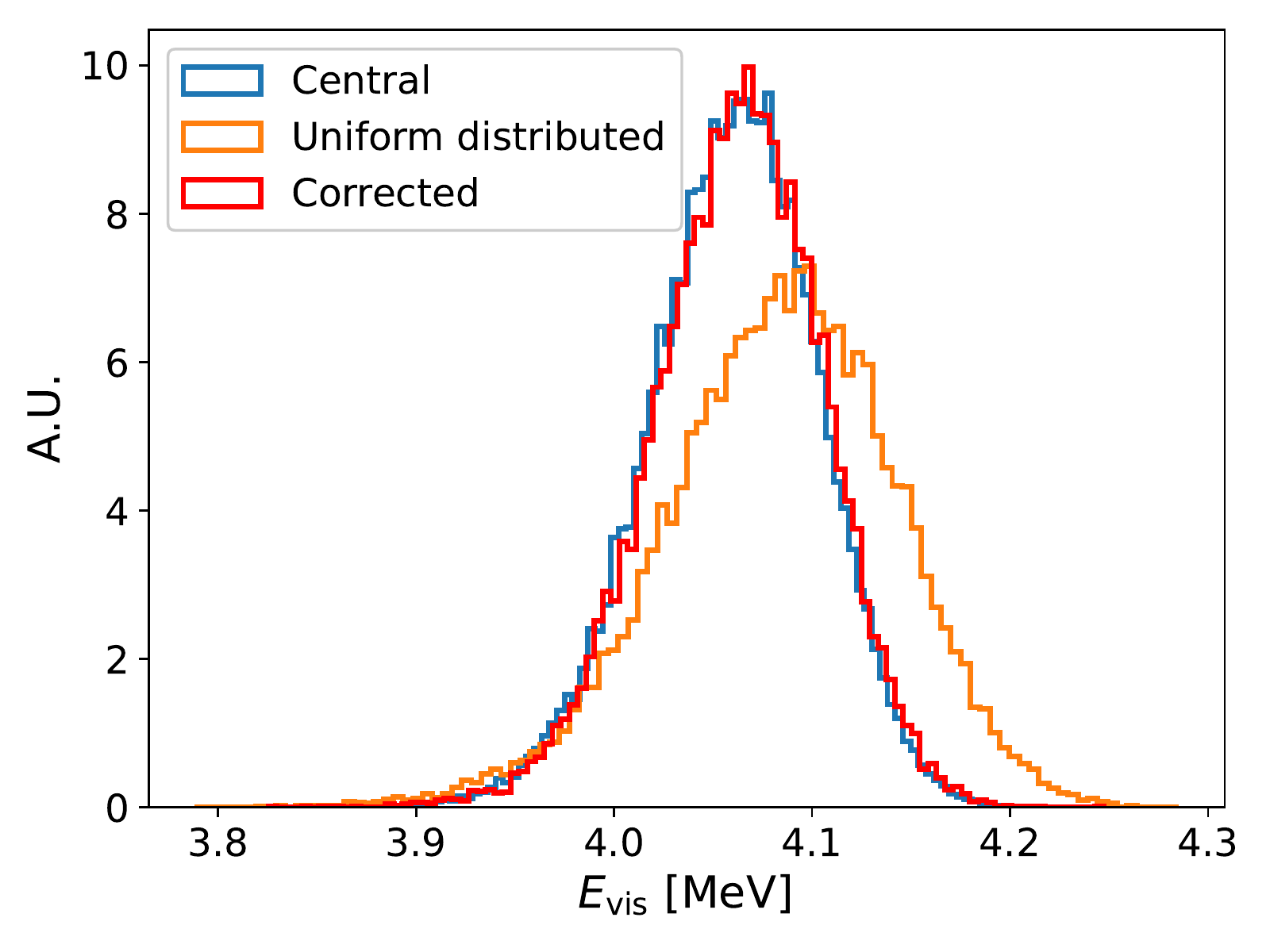}
\label{fig:reconstructed_spectrum}
}
\caption{(a) 3D surface of $g_{\rm ideal}(r,\theta)$ obtained by simulation of electrons with kinetic energy of \SI{1}{MeV} at many vertices in the detector. (b) $E_{\rm vis}$ spectrum of positron with kinetic energy of \SI{3}{MeV}. Blue line is the $E_{\rm vis}$ spectrum of positron at detector center, orange line is the $E_{\rm vis}$ spectrum of uniformly distributed positron, red line is the $E_{\rm vis}$ spectrum reconstructed by $g_{\rm ideal}(r,\theta)$ for uniformly distributed positron.}
\end{figure}

In order to correct the position dependence, we should use a non-uniformity map. To obtain the $g(r,\theta,\phi)$, electrons with kinetic energy of \SI{1}{MeV} are simulated at many vertices. The electron source is nearly a point source because it deposits energy in an area with a radius of a few millimeters. Since the detector is designed with rotational symmetry, we obtain $g(r,\theta,\phi) \approx g(r,\theta) $. The full non-uniformity map in ~Figure~\ref{fig:ideal_numap} can be obtained by Clough-Tocher two-dimension interpolator \cite{alfeld1984trivariate,nielson1983method,renka1984triangle}. We refer to this non-uniformity map as $g_{\rm ideal}(r,\theta)$.

Simulation shows that the $g_{\rm ideal}(r,\theta)$ can correct the non-uniformity of IBD positron well. We simulate the IBD events uniformly distributed within the CD and a set of the events in the center, then reconstruct uniformly distributed events by Equation~\ref{eq:reconstruct}. Figure~\ref{fig:reconstructed_spectrum} is an example for positrons with kinetic energy of \SI{3}{MeV}. The difference between the reconstructed positron energy spectrum and the center positron energy spectrum is small. Specifically, the relative difference between their central values and the relative energy resolution difference ($\Delta \sigma/E$) are less than 0.15\% and 0.01\%, respectively. This conclusion is applicable to the positrons in the kinetic energy range from 0 to \SI{8}{MeV}.

\subsection{Optimizing finite-point uniformity calibration}
In this section, we optimize the layout of the CLS system and select some deployment calibration points to obtain a good approximation of $g_{\rm ideal}(r,\theta)$. The optimization results also feed back into the design of the calibration system.
\subsubsection{Optimize anchor positions}
As shown in Figure~\ref{fig:ideal_numap},  $g_{\rm ideal}(r,\theta)$ is almost symmetrical about the plane of $\theta=90^{\circ}$. 
To be more specific, $|g_{\rm ideal}(r,\theta) - g_{\rm ideal}(r,180^{\circ} - \theta)|/g_{\rm ideal}(r,\theta)$ is less than $0.2\% $ within the fiducial volume thanks to the symmetrical arrangement of SiPMs and the symmetry of acrylic vessel. A minor asymmetry comes from different size of two holes at two poles of the acrylic vessel.  It is safe to assume $g(r,\theta) = g(r,180^{\circ} -  \theta)$.

When two anchors are fixed, we select very dense points $\vec{P}_{\rm calib}$ that can be reached by calibration system, and calculate $g_{\rm ideal}(\vec{P}_{\rm calib})$. Then, Clough-Tocher two-dimension interpolator is applied to get $g_{\rm calib}(r,\theta; \theta_{1},\theta_{2},\phi_{2})$, where $\theta_{1}$, $\theta_{2}$ and $\phi_{2}$ represent the positions of two anchors. Since we assume that the detector is symmetric about the z axis, we set $\phi_{\rm 1} = 0$ where $\phi_{\rm 1}$ is the azimuthal angle of one of the two anchors. In order to optimize the position of anchors, a penalty function 
\begin{small}
\begin{equation}
    L(\theta_{1},\theta_{2},\phi_{2}) = \int\limits_{S_{R}}(g_{\rm ideal}(r,\theta) - g_{\rm calib}(r,\theta; \theta_{1},\theta_{2},\phi_{2}))^2\cdot dV
\end{equation}
\end{small}
is defined to evaluate the difference between $g_{\rm ideal}(r,\theta)$ and $g_{\rm calib}(r,\theta; \theta_{1},\theta_{2},\phi_{2})$, where $S_{R}$ means a spherical volume whose radius is smaller than $R$. $R$ is set to \SI{700}{mm} to include fiducial volumes with a radius less than 650~mm. Besides, limitations $\theta_1 > 90^{\circ}$ and $\theta_2 > \theta_1$ are added to make installation and positioning of anchors more convenient. We obtain $\theta_{1} \approx 102.5^{\circ}$,\, $\theta_{2} \approx 155.2^{\circ}$,\, $\phi_{2} \approx 151.7^{\circ}$ by minimizing $L(\theta_1,\theta_2,\phi_2)$.

\subsubsection{Determine suitable calibration points}
Once the positions of two anchors are determined,  the position of the CLS cable is fixed. At true calibration, only limited calibration points along the fixed CLS cable can be used. The criterion to select suitable calibration points is putting more calibration points where the modulus of the gradient of $g(r,\theta)$ is large.
\paragraph{ACU} For ACU system, the rule is to set a calibration point every \SI{100}{mm} in the area with radius less than or equal to \SI{500}{mm}, and set a calibration point every \SI{50}{mm} in the area with a radius of \SI{500}{mm} to \SI{850}{mm}.
\paragraph{CLS} For CLS system, the rule is to start from the starting point of the CLS cable ($\vec P_0$), and add a calibration point ($\vec P_i$) when conditions $0.01 \leq |g_{\rm ideal}(\vec P_{i}) - g_{\rm ideal}(\vec P_{i-1})|$ and \SI{25}{mm} $\leq|\vec P_i - \vec P_{i-1}|$ are met at the same time, or when condition \SI{50}{mm} $\leq |\vec P_i - \vec P_{i-1}|$ is met. The starting point $\vec P_0$ is marked in Figure~\ref{fig:detector}.

\begin{figure}[htbp!]
\centering
\subfloat[]{
\includegraphics[width=0.48\textwidth]{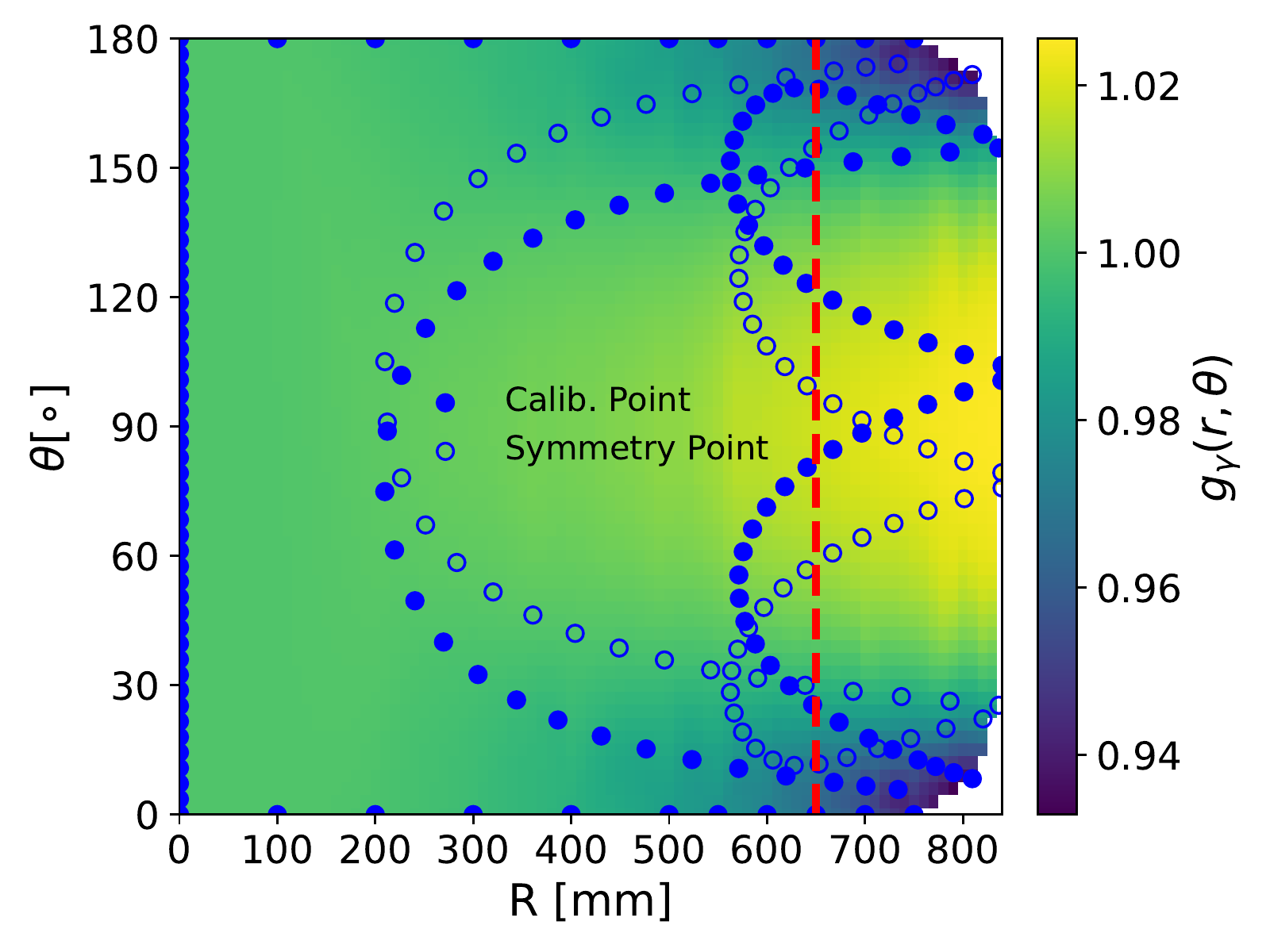}
\label{fig:calib_numap}
}
\hfill
\subfloat[]{
\includegraphics[width=0.48\textwidth]{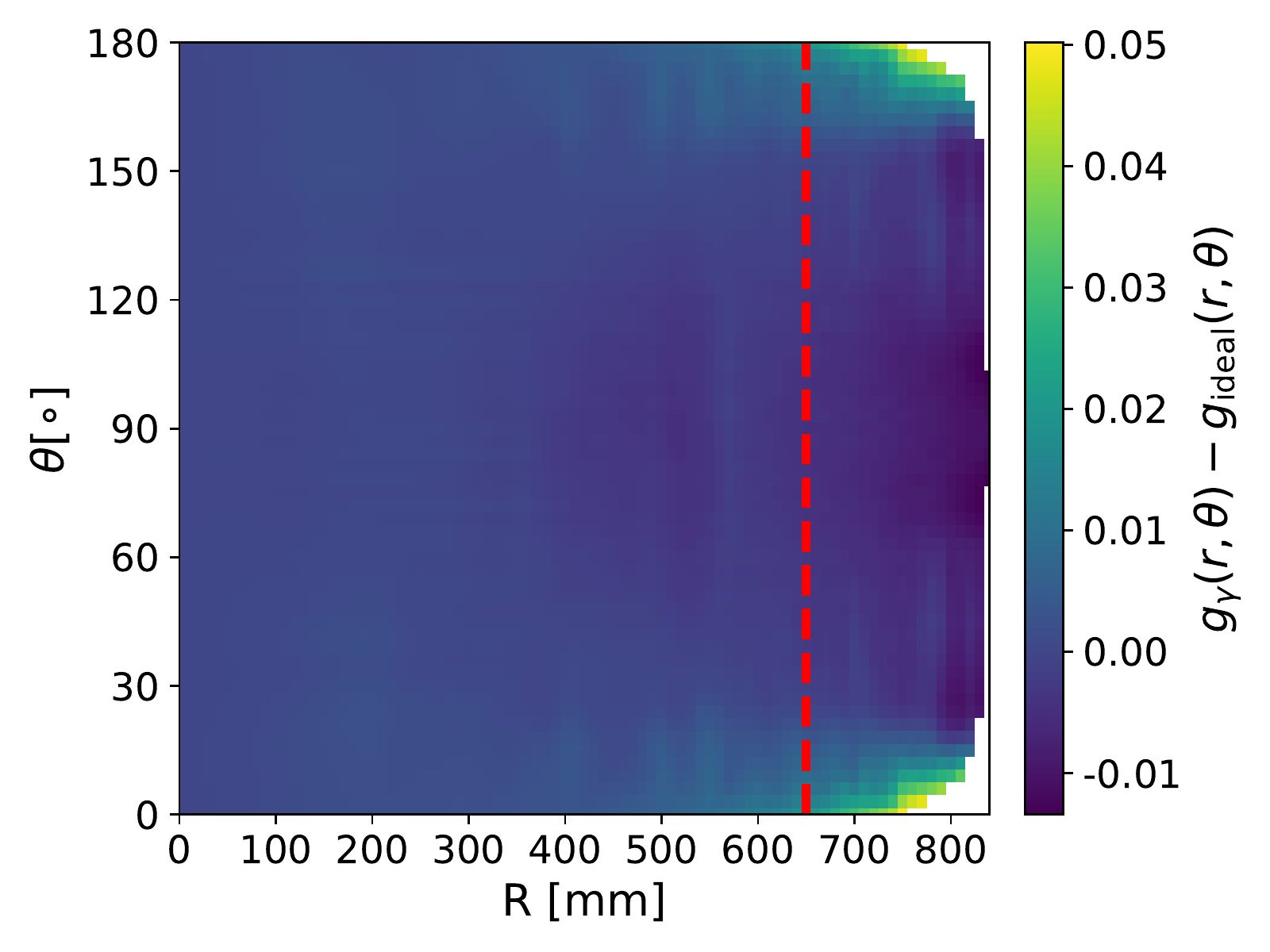}
\label{fig:calib_ideal_diff}
}
\caption{(a) $g_{\gamma}(r,\theta)$ interpolated from the $^{68}\rm Ge$ and $^{137}\rm Cs$ calibration data. Solid circles are real calibration points. Hollow circles are the symmetry points of the solid circles with respect to $\rm z=0$ plane. (b) $g_{\gamma}(r,\theta) - g_{\rm ideal}(r,\theta)$. Red line is the boundary of the fiducial volume.}
\end{figure}

Totally, 110 points for non-uniformity calibration are selected shown as the solid points in Figure~\ref{fig:calib_numap}.

\subsection{Verification of the calibration}
\label{sec:verify_nonuni_calib}

When performing non-uniformity calibration, we plan to use $^{68}$Ge and $^{137}$Cs on the ACU and use $^{137}$Cs on the CLS. We can not use $^{68}$Ge at CLS because visible energy of gammas produced by positron annihilation is mixed with the visible energy of positron kinetic energy. The main branch of $^{137} \rm Cs$ decay is through beta decay to the excited state of barium which then de-excites and emits gamma of 0.662 MeV. The mean half-life of the excited $^{137}\rm Ba$ is about 2.55 minutes and the end-point kinetic energy of the $\beta$ rays are about \SI{0.518}{MeV}, so the $\beta$ rays can be distinguished from that of 0.662 MeV gamma rays.

For a given calibration points on CLS, we place the $^{137}\rm Cs$ at there and obtain the photo-electron yield, then divide it by the photo-electron yield of $^{137}\rm Cs$ at the center. For a given calibration points on ACU, we place the $^{68}\rm Ge$ there and obtain the photo-electron yield, then divide it by the photo-electron yield of $^{68}\rm Ge$ at the center. Then we use the Clough-Tocher two-dimension interpolator to obtain the $g(r,\theta)$, as shown in Figure~\ref{fig:calib_numap}. Since this non-uniformity map is obtained using gamma sources, it is marked as $g_{\gamma}(r,\theta)$. Near two poles of the acrylic vessel, the fully absorbed energy peaks can not be fitted due to large energy loss effect. The difference between $g_{\gamma}(r,\theta)$ and $g_{\rm ideal}(r,\theta)$ is shown in Figure~\ref{fig:calib_ideal_diff}. This difference prevents us from perfectly correct the effects of non-uniformity, which means there is residual non-uniformity. The residual non-uniformity, denoted as RN, is defined as

\begin{gather}
    V_{\rm FV} = \int_{\rm FV} dV,\\
	\overline{\Delta g} = \frac{1}{V_{\rm FV}}\cdot \int_{\rm FV}(g_{\gamma}(r,\theta) - g_{\rm ideal}(r,\theta))\cdot dV,\\
	\rm RN = \sqrt{\frac{1}{V_{\rm FV}}\cdot \int_{\rm FV} (g_{\gamma}(r,\theta) - g_{\rm ideal}(r,\theta) - \overline{\Delta g})^{2}\cdot dV,}
\end{gather}
where FV is the fiducial volume. $\overline{\Delta g}$ is about -0.14\% and will cause bias of reconstructed visible energy. RN leads to degradation in the resolution of reconstructed visible energy, but it's only about 0.2\%, satisfying the requirement of less than 0.5\%. Thus, the energy non-uniformity can be well calibrated with the radioactive sources on ACU and CLS.

\subsection{Energy resolution}
\label{sec:energy_resolution}
\begin{figure}[]
\centering
\subfloat[]{
\includegraphics[width=0.48\textwidth]{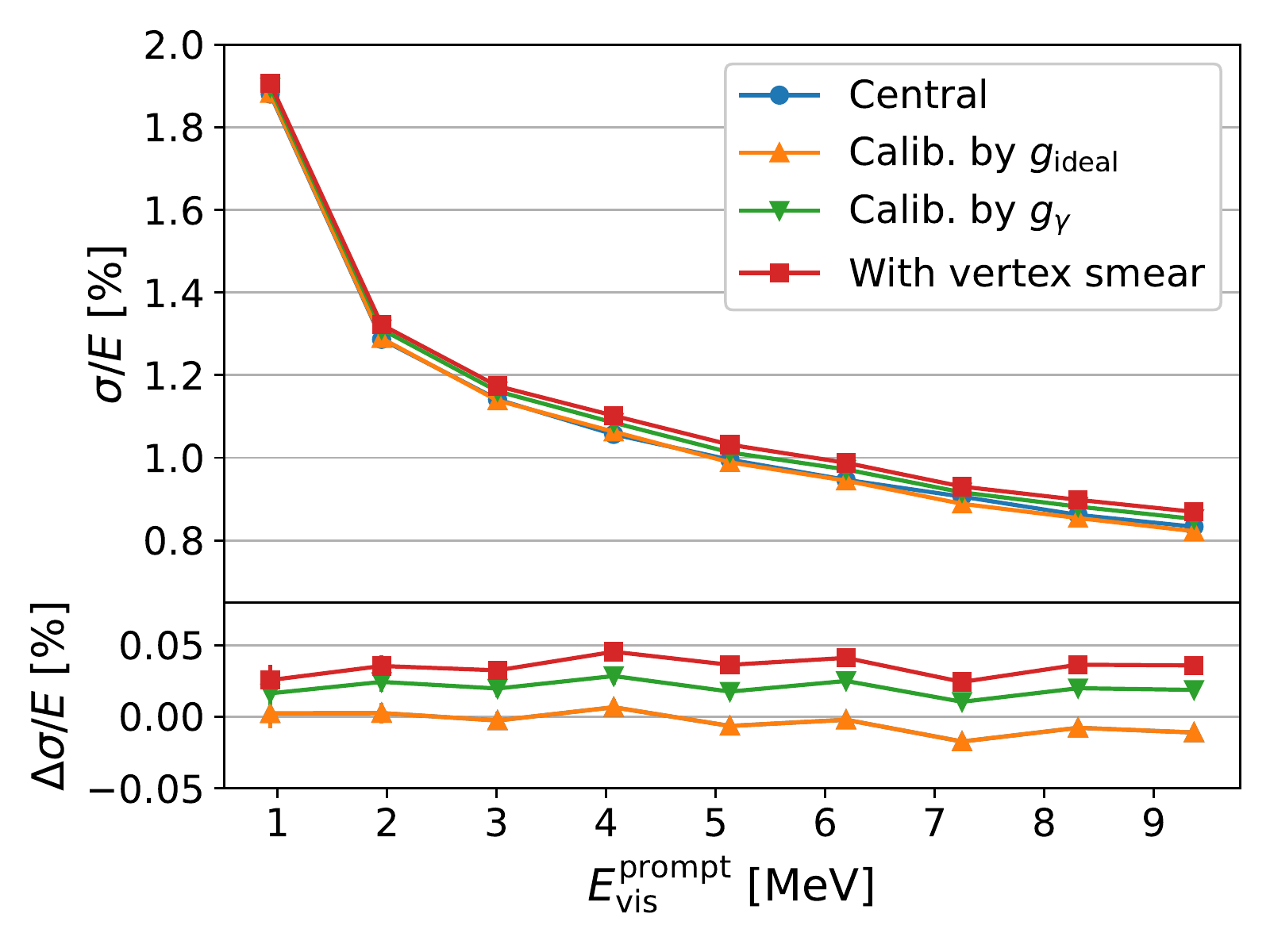}
\label{fig:IBD_rec_res}
}
\hfill
\subfloat[]{
\includegraphics[width=0.48\textwidth]{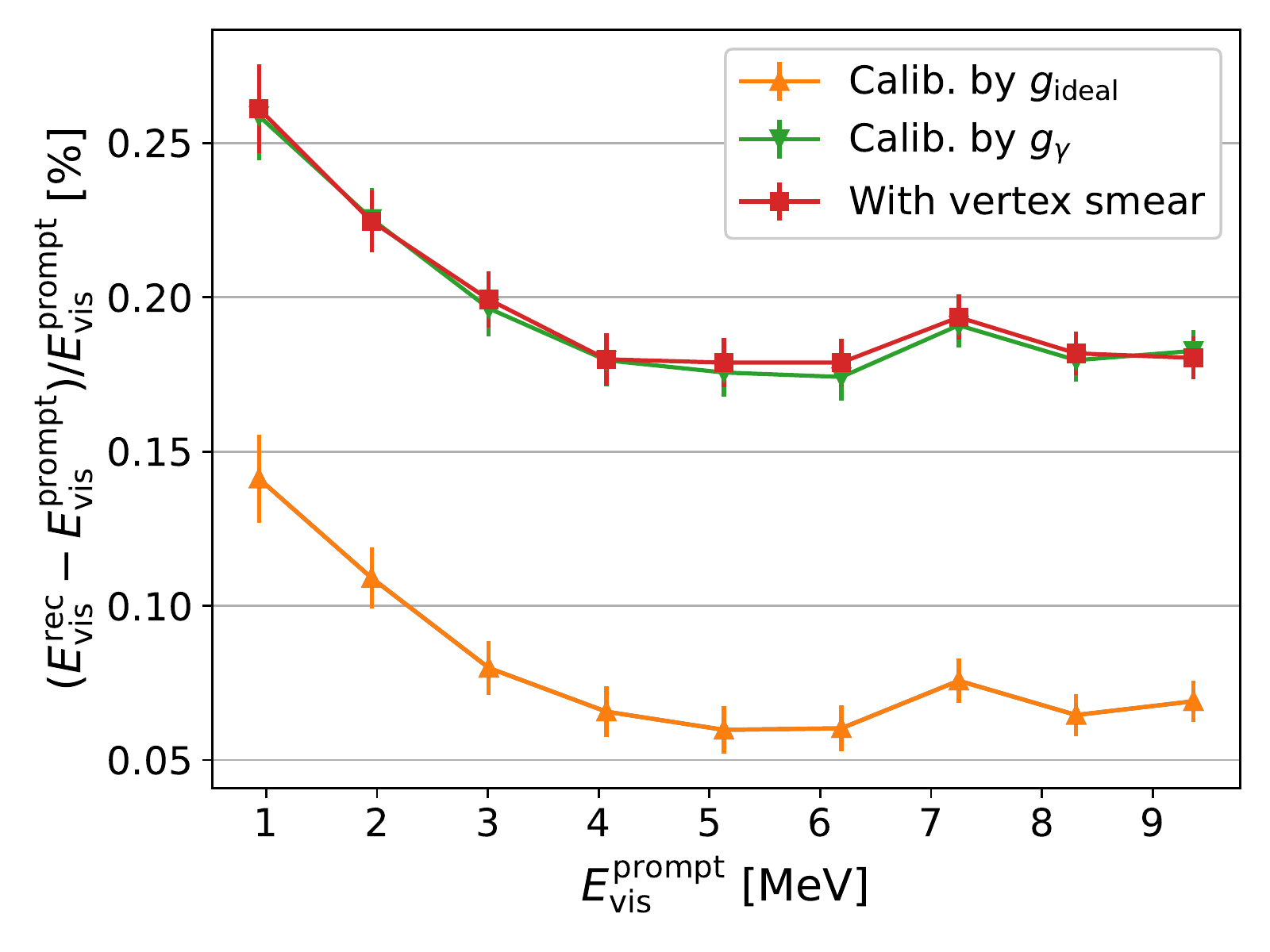}
\label{fig:IBD_rec_bias}
}
\caption{(a) Energy resolution of reconstructed IBD prompt events. (b) $E_{\rm vis}$ bias of reconstructed $e^{+}$ spectrum. $E^{\rm prompt}_{\rm vis}$ here means visible energy of central IBD prompt events. $E^{\rm rec}_{\rm vis}$ is the visible energy of reconstructed IBD prompt events which are distributed uniformly within the fiducial volume.}
\end{figure}

In order to get the energy resolution of the TAO, IBD events are simulated at the center of the CD. For mono-energetic IBD events, we can calculate the standard deviation ($\sigma$) and mean of their visible energies ($E^{\rm prompt}_{\rm vis}$). The energy resolution of IBD events is defined as $\sigma / E^{\rm prompt}_{\rm vis}$ and is shown in Figure~\ref{fig:IBD_rec_res}. TAO is able to achieve such unprecedented energy resolution for three main reasons. First, 95\% of the surface of the CD is covered with SiPMs with about 50\% photon detection efficiency. Second, the CD is small, so the liquid scintillation in the CD absorbs only a very small number of photons. Third, TAO operates in a low-temperature environment, which increases the photon yield~\cite{xie2021liquid}.

Residual non-uniformity and electronic effects such as cross talk, dark noise, charge resolution can cause the energy resolution degradation ($\Delta \sigma / E^{\rm prompt}_{\rm vis}$). The energy resolution degradation due to electronic effects is less than $0.23\%/E^{\rm prompt}_{\rm vis}$, see Ref.~\cite{tao_CDR} for details. To study the energy resolution degradation due to residual non-uniformity, Equation~\ref{eq:reconstruct} and $g_{\gamma}(r,\theta)$ are used to reconstruct uniformly distributed IBD events. During the reconstruction, a conservative vertex resolution of \SI{5}{cm} has been considered. The reconstructed visible energy spectrum of the IBD events whose reconstructed vertices are within FV is compared with the visible energy spectrum of the central IBD events. It can be seen from Figure~\ref{fig:IBD_rec_res} that the energy resolution degradation is less than 0.05\%. Besides, Figure~\ref{fig:IBD_rec_bias} shows that the bias of reconstructed energy is smaller than 0.3\%.

\section{Conclusion}
\label{sec:conclusion}
  A calibration strategy for the TAO detector has been developed to understand its physics non-linearity and non-uniformity. The TAO detector contains two independent calibration systems called the ACU and CLS. The ACU is capable to carry several radioactive and non-radioactive calibration sources and deploy one of them into the detector along the central vertical axis at each time, while the CLS is designed with a single radioactive source, that can be deployed to off-axis positions. For physics non-linearity, we utilize several gamma sources with energies ranging from a few hundred keV to several MeV to control it within 0.6\% for electron or positron with kinetic energy greater than \SI{0.5}{MeV}. $^{\rm 12}$B events can be used to reduce physics non-linearity down to 0.4\% assuming statistics collected in three years of data taking. For non-uniformity, we can utilize the ACU and CLS to deploy radioactive source to 110 positions to study the detector response, then generate a map to correct the detector non-uniformity. After the correction, residual non-uniformity is less than 0.2\%. The energy resolution degradation and energy bias caused by the residual non-uniformity can be controlled within 0.05\% and 0.3\% respectively. With this calibration strategy, TAO is able to measure high-precision reactor antineutrino energy spectrum.

\section*{Acknowledgement}
This work is supported by the National Natural Science Foundation of China under Grant Number 12022505, 11875282 and 11775247, the joint RSF-NSFC project under Grant Number 12061131008, the Youth Innovation Promotion Association of Chinese Academy of Sciences, and by the Program of the Ministry of Education and Science of the Russian Federation for higher education establishments with project Number FZWG-2020-0032 (2019-1569).

We gratefully acknowledge help from Jianglai Liu, Yue Meng, Feiyang Zhang, Zeyuan Yu and Miao Yu.

\appendix
\section{Ultraviolet LED calibration system}
\begin{figure*}[ht]
\centering
\includegraphics[width=\textwidth]{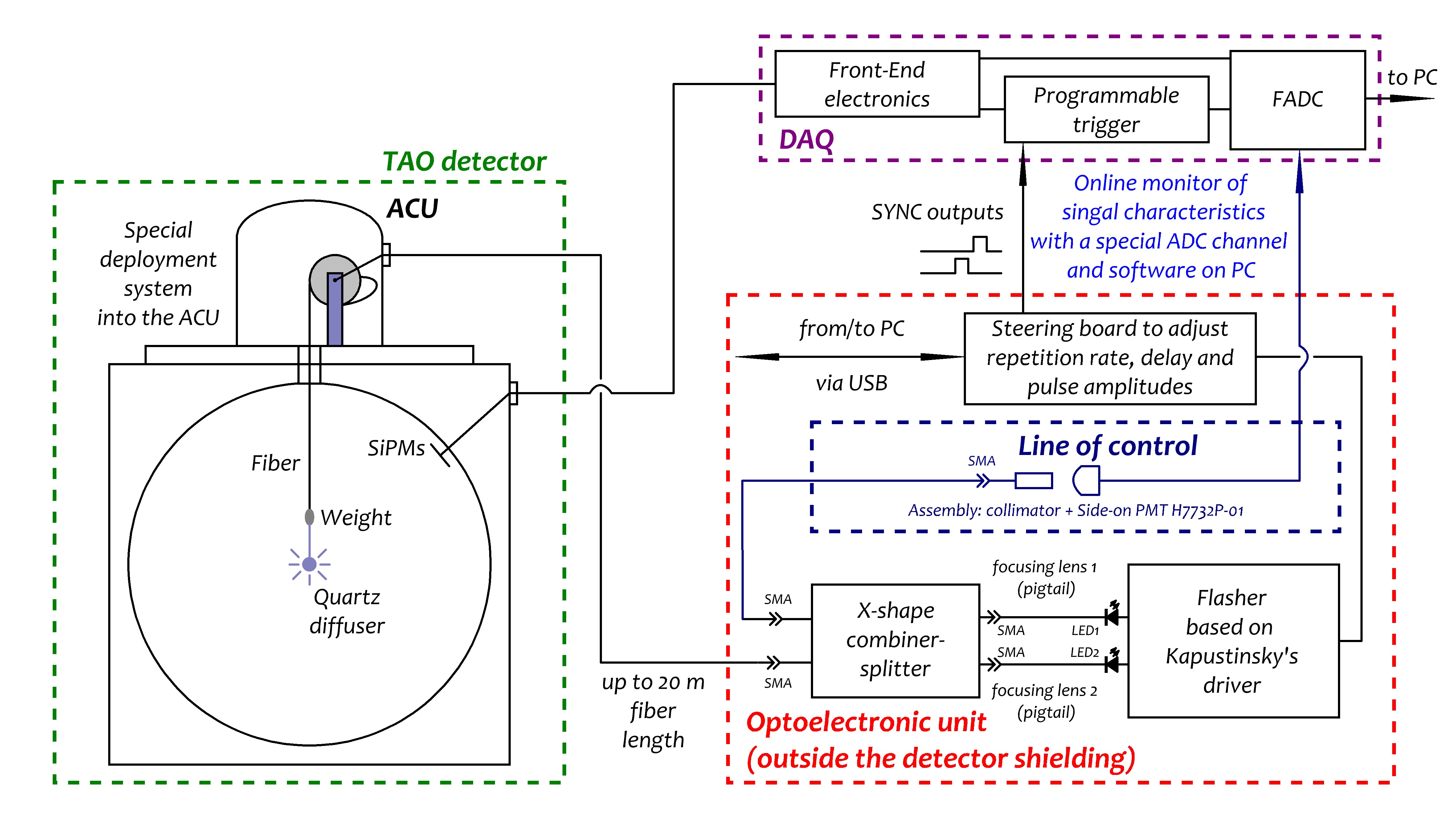}
\caption{The simplified scheme of the UV LED calibration subsystem. The dashed green, purple and red boxes show structural parts of the experimental setup namely the detector, the DAQ electronics and the optoelectronic unit of the UV LED subsystem. The last one generates light pulses and sends them to the detector target via a fiber. Simultaneously, the DAQ trigger receives the relevant external triggers from the steering board of the UV LED subsystem and the ADC gets signals from the control line. See more details in the text.}
\label{fig:UVLED_calib}
\end{figure*}

The UV source is a part of the UV LED calibration subsystem. As shown in Figure~\ref{fig:UVLED_calib}, the system comprises the optoelectronic unit placed out of the detector shielding, a deployment apparatus inside the ACU and a long optical fiber with a special diffuser at its tip, to improve the isotropy of radiation. The UV light is transmitted to the CD via the fiber and the deployment device inserts the fiber and diffuser into the detector. The wavelength is $(265 \pm 5)$~\SI{}{nm} by default and can be changed to \SI{420}{nm} (blue) or any other value  if necessary.

The UV LED calibration subsystem is designed to achieve three main goals. The first one is monitoring the stability of the TAO detector parameters. This task includes monitoring the health of individual channels and calibrate their timing, SiPM gain, quantum efficiency (QE). 

The second goal is imitation of real physical processes to test the Data Acquisition (DAQ) and the offline analysis pipeline. This includes simulations of the IBD events and background events, and tests on the high rate and low energy threshold data acquisitions. A study of pileup in the CD is the last main goal.

Such a wide use of the UV LED calibration subsystem is possible due to its specific design. The driver of the nanosecond LED pulser (flasher) is made according to Kapustinsky’s  basic  scheme~\cite{Kapustinsky:1985hnc}. There are two LED channels in the flasher that can serve to generate two consecutive signals or increase amplitude of a single output signal by merging the first and second signals with each other. Moreover, both LEDs work independently and the respective signals can have different amplitudes. This flexibility is provided by a steering board (so-called imitator board) with a microcontroller. It also allows to adjust the repetition rate. All the settings can be configured remotely using a computer with a USB Virtual COM Port. The output signals are monitored pulse to pulse with a control line. To do this, we merge the signals into a single time sequence and then make two identical copies of the series using an X-shape combiner-splitter. One of them goes to the detector and the other is sent to the control line. In fact, the second signal is registered with a side-on Photo-multiplier Tube (PMT) (Hamamatsu H7732P-01 in this case) which is connected to the TAO DAQ. The simplified scheme of the UV LED calibration subsystem is shown in Figure~\ref{fig:UVLED_calib}. The presented design is the development of the concept proposed in references~\cite{Chepurnov:2016hdg,Chepurnov:2017ivh} and realized with some changes in the JUNO Laser Calibration System~\cite{Zhang:2018yso}.

\bibliography{calibration}
\end{document}